\documentclass[aps,prl,twocolumn,reprint,amsmath,amssymb,preprintnumbers,superscriptaddress,bibnotes,nofootinbib,floatfix]{revtex4-1}

\usepackage{graphicx}
\usepackage{dcolumn}
\usepackage{bm}
\usepackage{color}
\usepackage{amssymb}
\usepackage{amsmath}
\usepackage[colorlinks=true,allcolors=Purple,pdfusetitle]{hyperref}
\usepackage{rotating}
\usepackage{multirow}
\usepackage{graphicx}
\usepackage[dvipsnames]{xcolor}
\usepackage{svg}
\usepackage{esdiff}
\definecolor{Green}{RGB}{0, 128, 0}
\newcommand{\orcid}[1]{\href{https://orcid.org/#1}{\includegraphics[width=10pt]{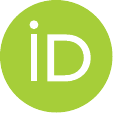}}}
\usepackage{CJK}

\newcommand{\prlsection}[2]{{\it\textbf{#1}{#2}}---}
\makeatletter

\begin{document}

\preprint{CETUP2024-009}
\preprint{FERMILAB-PUB-24-1033-T}
\preprint{INT-PUB-24-050}
\preprint{N3AS-24-033}

\title{Non-conservation of Lepton Numbers in the Neutrino Sector \\ Could Change the Prospects for Core Collapse Supernova Explosions}

\author{Anna M. Suliga \orcid{0000-0002-8354-012X}}
\email{a.suliga@nyu.edu}
\affiliation{
Department of Physics, University of California Berkeley, Berkeley, California 94720, USA}
\affiliation{
Department of Physics, University of Wisconsin--Madison,
Madison, Wisconsin 53706, USA}
\affiliation{
Department of Physics, University of California, San Diego, La Jolla, CA 92093-0319, USA}
\affiliation{Center for Cosmology and Particle Physics, New York University, New York, NY 10003, USA}

\author{Patrick Chi-Kit Cheong \begin{CJK*}{UTF8}{bkai}(張志杰)\end{CJK*} \orcid{0000-0003-1449-3363}}
\email{patrick.cheong@berkeley.edu}
\affiliation{
Department of Physics, University of California Berkeley, Berkeley, California 94720, USA}
\affiliation{
Department of Physics \& Astronomy, University of New Hampshire, 9 Library Way, Durham NH 03824, USA}

\author{Julien Froustey \orcid{0000-0002-6466-8232}}
\email{jfroustey@berkeley.edu}
\affiliation{
Department of Physics, University of California Berkeley, Berkeley, California 94720, USA}
\affiliation{
Department of Physics, University of California, San Diego, La Jolla, CA 92093-0319, USA}
\affiliation{
Department of Physics, North Carolina State University, Raleigh, North Carolina 27695, USA}

\author{George M.~Fuller \orcid{0000-0002-9716-9552}}
\email{gfuller@physics.ucsd.edu}
\affiliation{
Department of Physics, University of California, San Diego, La Jolla, CA 92093-0319, USA}

\author{{Luk\'{a}\v{s} Gr\'{a}f} \orcid{0000-0001-6181-9210}}
\email{lukas.graf@nikhef.nl}
\affiliation{
Department of Physics, University of California Berkeley, Berkeley, California 94720, USA}
\affiliation{
Department of Physics, University of California, San Diego, La Jolla, CA 92093-0319, USA}
\affiliation{
Nikhef, Theory Group, Science Park 105, 1098 XG Amsterdam, The Netherlands}
\affiliation{
Institute of Particle and Nuclear Physics, Faculty of Mathematics and Physics, Charles University in Prague, V Hole\v{s}ovi\v{c}k\'ach 2, 180 00 Praha 8, Czech Republic}

\author{Kyle Kehrer \orcid{0000-0002-8714-1599}}
\email{kkehrer@ucsd.edu}
\affiliation{
Department of Physics, University of California, San Diego, La Jolla, CA 92093-0319, USA}

\author{Oliver Scholer \orcid{0000-0002-0337-1486}}
\email{scholer@mpi-hd.mpg.de}
\affiliation{
Max-Planck-Institut f\"{u}r Kernphysik, Saupfercheckweg 1, 69117 Heidelberg, Germany}

\author{Shashank Shalgar \orcid{0000-0002-2937-6525}}
\email{shashank.shalgar@nbi.ku.dk}
\affiliation{
Niels Bohr International Academy and DARK, Niels Bohr Institute,
University of Copenhagen, Blegdamsvej 17, 2100, Copenhagen, Denmark}

\date{June 19, 2025}


\begin{abstract}
We show that interactions violating the conservation of lepton numbers in the neutrino sector could significantly alter the standard low entropy picture for the pre-supernova collapsing core of a massive star.
A rapid neutrino-antineutrino equilibration leads to entropy generation and enhanced electron capture and, hence, a lower electron fraction than in the standard model. This would affect the downstream core evolution, the prospects for a supernova explosion, and the emergent neutrino signal. If realized by lepton-number-violating neutrino self-interactions (LNV $\nu$SI), the relevant mediator mass and coupling ranges can be probed by future accelerator-based experiments.
\end{abstract}

\maketitle

\prlsection{Introduction}{.}
Core-collapse supernovae (CCSN) are not only fundamental to our understanding of much of astrophysics and cosmology, but they are in essence weak interaction phenomena and, as such, are alluring venues for studying beyond standard model (BSM) physics, especially in the neutrino sector~\cite{Raffelt:1996wa}. In this letter we investigate the effects of lepton number-violating (LNV) physics on the early stages of the collapse of supernova progenitor stars, i.e., those with initial masses in excess of $\sim 10\,{\rm M}_\odot$. These stars will suffer prodigious neutrino energy and entropy losses during their short evolutionary time up to the point of core instability and collapse. This entropy loss renders the cores of these stars thermodynamically \lq\lq cold,\rq\rq\ with low entropy-per-baryon, $s_{k_\mathrm{B}} \sim 1$ in units of Boltzmann's constant $k_{\rm B}$, with the support pressure dominated by relativistically degenerate electrons. That, in turn, sets up these configurations for instability and collapse. Our study exploits the sensitivity of the subsequent 
collapse-phase physics to LNV neutrino interactions. This sensitivity arises because such processes can tap into the huge zero point energy reservoir of degenerate electron lepton number ($\sim {10}^{57}$) that characterizes the collapsing, pre-bounce core.  

The framework for how the weak interaction dictates CCSN core evolution and collapse to a neutron star was confirmed in broad brush by SN1987A~\cite{Kamiokande-II:1987idp, Bionta:1987qt, ALEXEYEV1988209}. The inferred neutrino luminosity was consistent with most of the gravitational binding energy of the proto-neutron star carried off by $\sim {10}^{58}$ neutrinos within $\sim 10\;{\rm s}$~\cite{Burrows:1986me}. The huge neutrino flux expected from CCSN can serve as a laboratory for neutrino physics~\cite{Zatsepin:1968kt, Loredo:2001rx, Pagliaroli:2010ik, Nardi:2003pr, Nardi:2004zg, Beacom:2000ng, Lu:2014zma, Rossi-Torres:2015rla, Hansen:2019giq, Pompa:2022cxc, Pitik:2022jjh}, the vexing problem of medium-affected nonlinear neutrino flavor oscillations~\cite{Duan:2010bg, Chakraborty:2016yeg, Tamborra:2020cul, Patwardhan:2021rej,Richers:2022zug, Volpe:2023met} and key neutrino mass physics issues~\cite{Dighe:1999bi, KATRIN:2021uub, Katrin:2024tvg}. 
The high densities and large electron lepton numbers associated with the CCSN collapse phase preclude neutrino flavor transformation~\cite{1987ApJ...322..795F}, but the downstream post-bounce physics may be affected by these. 
Nevertheless, the experimental discovery of neutrino oscillations is direct evidence for BSM physics~\cite{Super-Kamiokande:1998kpq, Hirata:1992ku, Bellerive:2016byv}, spurring proposals for novel experimental and astrophysical probes.

Non-standard neutrino physics could influence neutrino decoupling and Big Bang Nucleosynthesis (BBN) in the early universe. The fossil record of these processes as observed indirectly in the cosmic microwave background (CMB) and cosmic neutrino background (C$\nu$B), e.g., via large scale structure considerations, could provide BSM physics discovery or constraint channels~\cite{Beacom:2004yd, Bell:2005dr, Grohs:2016cuu, Grohs:2020xxd, Froustey:2021azz, Froustey:2024mgf}. \lq\lq Secret\rq\rq\ neutrino self-interactions ($\nu$SI) as proposed in Ref.~\cite{Bialynicka-Birula:1964ddi} have been invoked as a way of coupling the neutrino fluid to itself at the epoch of CMB decoupling, e.g., enabling a potential solution to the Hubble tension problem~\cite{Oldengott:2017fhy, Lancaster:2017ksf, Park:2019ibn, Barenboim:2019tux, Blinov:2019gcj, Huang:2021dba}. However, these BSM interactions need not be LNV. The LNV processes we consider may make little difference in a standard, low-lepton-number early universe epoch for neutrino decoupling or BBN~\cite{Grohs:2020xxd}. 
CCSN may provide LNV $\nu$SI insights that are unique or complementary to those gleaned from the early universe. Several works have studied $\nu$SI in CCSN~\cite{Dicus:1982dk, Kolb:1987qy, Li:2017mfz, Shalgar:2019rqe, Reddy:2021rln, Mukhopadhyay:2021gox, Chang:2022aas, Bhattacharya:2023wzl} (see Refs.~\cite{Fiorillo:2023cas, Fiorillo:2023ytr} for a recent discussion), including some centered on Majoron-like~\cite{Chikashige:1980qk, Gelmini:1980re, Schechter:1981cv} models~\cite{Kolb:1981mc, Fuller:1988ega, Aharonov:1988ee, Choi:1987sd, Grifols:1988fg, Konoplich:1988mj, Berezhiani:1989za, Choi:1989hi, Farzan:2002wx, Rampp:2002kn, Akita:2022etk, Fiorillo:2022cdq, Akita:2023iwq, Telalovic:2024cot}, and others on $\nu$SI-induced
alterations of neutrino flavor transformation~\cite{Blennow:2008er}. The $\nu$SI can also modify the high-energy astrophysical neutrino flux~\cite{Ng:2014pca, Bustamante:2020mep, Creque-Sarbinowski:2020qhz, Esteban:2021tub}.

\prlsection{LNV in the Neutrino Sector}{.}
To assess impacts of the lepton-number-violating interactions on collapse let us adopt a $\nu$SI mediated by a scalar field $\phi$ with
\begin{equation}
\label{eq:Lagrangian_scalar_LNV}
    \mathcal{L}^{\phi} = g_{\phi,\alpha\iota} \, \phi \, \overline{\nu_{L,\alpha}} \, \nu_{L,\iota}^c \ ,
\end{equation}
where in general we can take this interaction to be LNV in one or more neutrino and antineutrino flavors $\alpha,\iota$. 
Here we will consider three $\nu_e$ conversion cases that could alter the standard model of stellar collapse: Case (1) $\nu_e \rightleftharpoons \nu_e,\bar\nu_e,\nu_\mu,\bar\nu_\mu,\nu_\tau,\bar\nu_\tau$, Case (2) $\nu_e \rightleftharpoons \nu_e,\bar\nu_e,\nu_x,\bar\nu_x$, and Case (3) $\nu_e\rightleftharpoons \nu_e,\bar\nu_e$. Note that in the heavy mediator limit Case (3) is suppressed for the Lagrangian in Eq.~\eqref{eq:Lagrangian_scalar_LNV}~\cite{Jenkins:2017jig, Fonseca:2018aav}.
For Cases (1) and (2), for mediator masses much larger than the momentum exchanged in the interaction, and in the perturbative coupling limit, neutrino-neutrino scattering cross sections can be approximated as~\cite{Kolb:1981mc, Fuller:1988ega, Esteban:2021tub}
\begin{equation}
\label{eq:cross-section}
\sigma_{\nu \mathrm{SI}} \approx \frac{G_{\nu\mathrm{SI}}^2}{8\pi} {E_{\nu}^{(1)} E_{\nu}^{(2)} (1 - \cos\theta)} \ ,
\end{equation}
where we define the effective neutrino self-coupling $G_{\nu\mathrm{SI}} = g_{\phi}^2/m_\phi^2$ with $g_\phi$ and $m_\phi$ standing for the coupling (assumed to be identical among all flavors) and mass of the new mediator, $E_\nu^{(1,2)}$ being the energies of the interacting neutrinos and $\cos\theta$ denoting the angle between them. For our considerations, the mediator mass must be $m_\phi \gtrsim 100\,$MeV to avoid on-shell production during the collapse phase.

\begin{figure}[t]
\includegraphics[width=0.98\columnwidth]{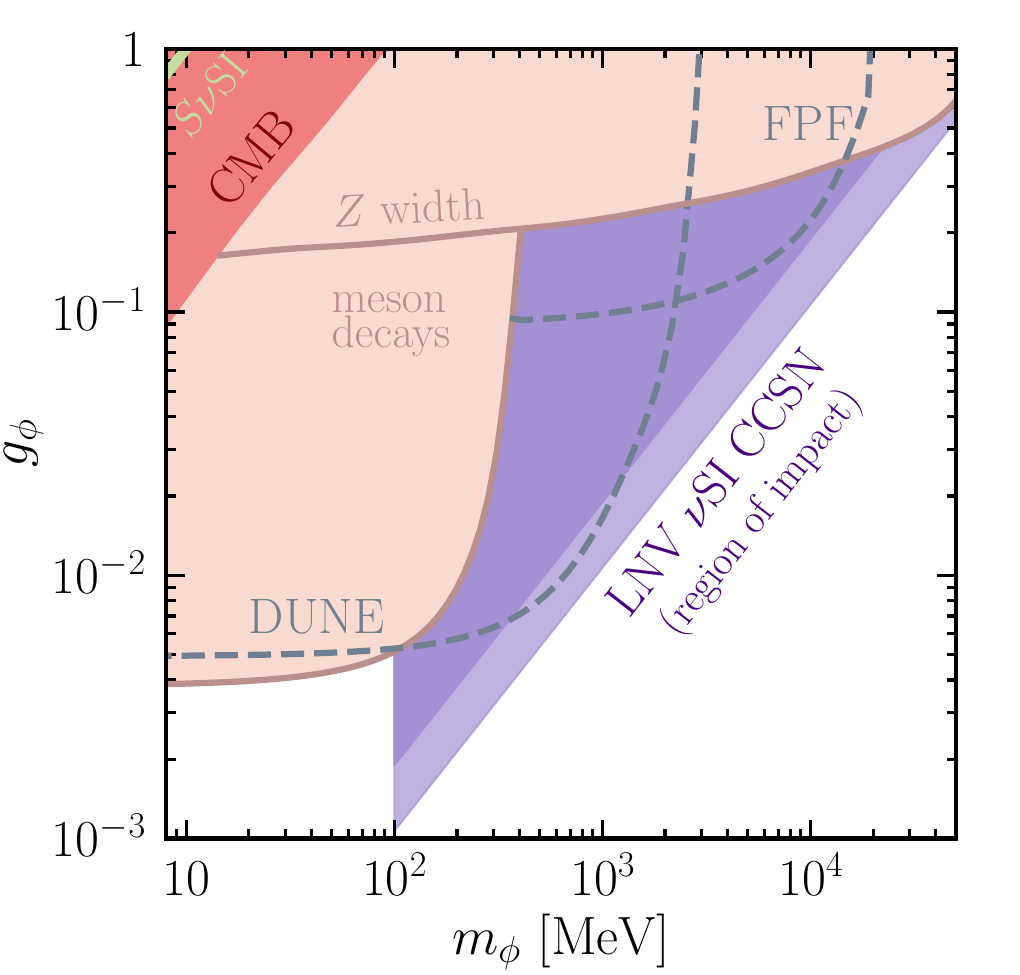}
\caption{The region of the LNV $\nu$SI parameter space studied in this work together with the terrestrial limits from the $Z$ width and meson decay measurements~\cite{Pang:1973rxr, Bilenky:1992xn, Laha:2013xua, Berryman:2018ogk}, cosmological impacts~\cite{Park:2019ibn, Archidiacono:2013dua}, and projected experimental sensitivities for DUNE~\cite{Berryman:2018ogk} and FPF~\cite{Kelly:2021mcd}. The dark and light violet shading shows parameters that will alter the standard CCSN collapse physics (see text).}
\label{Fig0}
\end{figure}

Fig.~\ref{Fig0} shows the experimental constraints on $\nu$SI and LNV $\nu$SI coupling $g_\phi$ and mediator mass $m_\phi$ that stem from meson decay and $Z$-width measurements~\cite{Pang:1973rxr, Bilenky:1992xn, Laha:2013xua, Pasquini:2015fjv, Berryman:2018ogk} (1-loop contributions may affect these~\cite{Dev:2024twk}).
The $\nu$SI parameters suggested at CMB decoupling to effect a Hubble tension fix~\cite{Park:2019ibn} are also shown (green band, upper left), along with parameters that affect or are impacted by CMB considerations (red band)~\cite{Archidiacono:2013dua}. 
The LNV $\nu$SI mediator mass and coupling ranges that most affect the collapse epoch are shown as violet shaded regions. The darker of these corresponds to our most conservative estimate of the LNV $\nu$SI strength required for these to dominate over the relevant weak interactions, while the lighter shaded region corresponds to a less stringent, but still plausible, criterion for this. Note that much of the LNV $\nu$SI parameter space relevant for significant collapse epoch effects can be probed by future accelerator-based experiments such as DUNE~\cite{Berryman:2018ogk, Ovchynnikov:2022rqj} and Forward Physics Facility (FPF)~\cite{Kelly:2021mcd}.

\prlsection{Standard core-collapse supernova evolution}{.} 

With Standard Model physics the lead up to collapse of the CCSN core is dominated by neutrino emission-driven entropy loss. At the onset of collapse the core central density is $\rho_{10} \sim 1$ (where $\rho_{10} \equiv \rho/{{10}^{10}\,{\rm g}\,{\rm cm}^{-3}}$), the temperature is $T \sim 1\,{\rm MeV}$, and the entropy-per-baryon is $s_{k_\mathrm{B}}\sim 1$~\cite{Bethe:1979zd}. The low entropy implies that the baryonic component is mostly contained in heavy nuclei, with free neutrons comprising only $\sim 10\%$ of the total mass fraction, and a much smaller free proton mass fraction, very sensitive to $T$. The core at this point is already significantly neutronized, with an electron fraction (number of electrons per baryon) of $Y_e \approx 0.4$~\cite{Bethe:1985sox}.
The relativistically degenerate electron component has a Fermi energy $\mu_e \approx 11.1\,{\rm MeV} {\left( \rho_{10}\, Y_e\right)}^{1/3}$. At the onset of collapse this is $\mu_e \approx 7\,{\rm MeV}$. Subsequently, as $\rho_{10}$ and hence $\mu_e$ increase during collapse, electron capture on both free protons and those in nuclei will drive $Y_e$ lower. 
In the early stages of collapse the $\nu_e$ produced by electron capture escape the core, so electron capture on free protons lowers the entropy. However, entropy can be increased by electron capture on protons \emph{inside} nuclei because these nuclei can be left in nuclear excited states~\cite{Bethe:1979zd, Fuller:1981mu}. 
When densities in the collapsing core reach $\rho_{10} \approx 10$ to $100$, the mean nuclear mass number $\langle A\rangle$ of the nuclei becomes so large that neutrino neutral current coherent scattering~\cite{Freedman:1973yd, COHERENT:2017ipa}, with cross section $\sigma_{\nu A} \propto {\langle A\rangle}^2\,E_\nu^2$, leads to neutrinos having mean free paths smaller than the core size, hence they are trapped. Weak (beta) equilibrium is attained quickly thereafter. In beta equilibrium the chemical potentials of all species are related by the Saha equation $\mu_e-\mu_{\nu_e} = {\hat{\mu}} + \delta m_{np}$, where the $\nu_e$ chemical potential is $\mu_{\nu_e}$, and ${\hat{\mu}}=\mu_n-\mu_p$ is the difference of neutron and proton kinetic chemical potentials, and $\delta m_{np}\approx 1.29\,{\rm MeV}$ is the neutron-proton mass difference.

In the approach to beta equilibrium, neutrino energies are redistributed not by conservative coherent scattering on nuclei, but rather by neutrino-electron scattering and by neutral current de-excitation of nuclei into neutrino pairs of all flavors~\cite{Fuller:1991kua, Misch:2013aq, Fischer:2013qpa, Misch:2016iwm, Fischer:2016boc, Dzhioev:2023sod}. The latter process and other thermal neutrino pair creation processes build up zero lepton number populations of $\nu_\mu$, $\bar\nu_\mu$, $\nu_\tau$, $\bar\nu_\tau$. Likewise, these create $\bar\nu_e$ as well, with the beta equilibrium condition dictating that the Fermi level of $\bar\nu_e$ is much lower than that of $\nu_e$.
At trapping the overall electron lepton number is $Y_{L_e}=Y_e+Y_{\nu_e} \sim 0.34$, roughly with $Y_e \approx 0.3$ and $Y_{\nu_e} \approx 0.04$, implying a $\nu_e$ Fermi level $\mu_{\nu_e} \approx 11.1\,{\rm MeV}\,{\left( 2 \rho_{10}\, Y_{\nu_e} \right)}^{1/3} \sim 20\,{\rm MeV}$. 

Subsequent to trapping, as the collapse proceeds, neutrino diffusion out of the core may lower the overall central electron lepton number at core bounce to $\mathcal{O}(0.25)$~\cite{2012ARNPS..62..407J, Bruenn:2012mj, Bruenn:2014qea, Lentz:2015nxa, 2012ARNPS..62..407J, Bruenn:2014qea, OConnor:2018sti, Janka:2016fox, Mezzacappa:2020oyq, Burrows:2020qrp}. Nevertheless, the large degenerate electron lepton number fraction contributes the bulk of the pressure, almost until the central density reaches nuclear density, $\rho_{10} \approx {10}^4$. The collapse is abruptly halted at this point, and a shock forms at the edge of the homologous core. The homologous core that serves as a piston for the shock is the inner, causally connected portion of the core inside the sonic point. As the shock moves out through the remainder of the core, the entropy jump $\Delta s_{k_\mathrm{B}} \sim 10$ outside the sonic point is sufficient to photo-dissociate the heavy nuclei. This costs $\sim 8\,{\rm MeV}$ per nucleon and, consequently, the shock evolves into a standing accretion shock. The shock is later revived by neutrino heating~\cite{Bethe:1985sox} facilitated by hydrodynamic transport~\cite{2012ARNPS..62..407J, Bruenn:2012mj, Bruenn:2014qea, Lentz:2015nxa, 2012ARNPS..62..407J, Bruenn:2014qea, Janka:2016fox, Mezzacappa:2020oyq, Burrows:2020qrp, Couch:2014kza,OConnor:2018tuw, Couch:2013coa, Takiwaki:2013cqa}.

\prlsection{Core-collapse supernova evolution with $\nu$SI}{.}
LNV $\nu$SI of sufficient strength would alter this standard picture.
In broad brush, at trapping when the neutrino number density rises, the LNV $\nu$SI could convert $\nu_e$ into neutrinos and antineutrinos of other flavors or types. 
This could reduce the $\nu_e$ Fermi level, altering weak equilibrium by opening phase space for electron capture with many downstream effects.
In what follows we present results for Case (1). Results for Cases (2) and (3) are discussed in the Supplemental Material.

For Case (1), $\nu_e \rightleftharpoons \nu_e,\bar\nu_e,\nu_\mu,\bar\nu_\mu,\nu_\tau,\bar\nu_\tau$, all neutrino flavors rapidly thermalize among themselves to a new \lq\lq neutrino temperature,\rq\rq\ $T_\nu$, larger than the matter temperature, $T_e$.
Only the ordinary charged current weak interactions, operating on a longer time scale, can bring the neutrinos and matter back into equilibrium with a common temperature. Re-establishment of equilibrium leads to entropy generation sufficient to shift the nuclear statistical equilibrium (NSE) characterizing the baryonic component from heavy nuclei to, possibly, all free nucleons and alpha particles. The lowering of $\mu_{\nu_e}$ opens holes for electron capture. This, combined with the larger number of free protons, leads to a precipitous drop in $Y_e$ and, hence, the electron pressure. The net result is that the core will be dominated by non-relativistic pressure. 

The key to this radically altered collapse history is that the $\nu$SI be much faster than the weak interactions other than the large coherent neutral current interactions. The latter trap the neutrinos, but have no role in mediating energy exchange between the neutrinos and matter. Therefore, requiring that the $\nu$SI interaction mean free path, $\lambda_{\nu{\rm SI}}$, be smaller than the core size, $R_\mathrm{C}$, should suffice as a fast $\nu$SI criterion. That condition is met when the $\nu$SI coupling strength is $G_{\nu\mathrm{SI}} \gtrsim (R_\mathrm{C}\, n_{\nu}\, \langle E_\nu \rangle^2/ 8 \pi)^{-1/2} \approx 3.6 \times 10^{-4}~\mathrm{GeV}^{-2}$ for the cross section from Eq.~\eqref{eq:cross-section}. The last estimate follows from taking $\lambda_{\nu{\rm SI}} \lesssim 10\;{\rm km}$ for $E_\nu = 10\;{\rm MeV}$ and $Y_{\nu_e} = 0.04$ at density $\rho_{10}=100$. This is a conservative condition since at trapping $R_\mathrm{C} \gg 10\,{\rm km}$. This estimate yields a coupling that is not excluded by laboratory experiments (see Fig.~\ref{Fig0}). For example, Kaon decay experiments provide limits at the level of approximately $G_{\nu\mathrm{SI}} \gtrsim 10^{-2}\;\mathrm{GeV}^{-2}$ around $m_\phi\approx 100\;\mathrm{MeV}$~\cite{Berryman:2022hds}.

\prlsection{Evolution toward a new weak equilibrium}{.} 
Once the LNV $\nu$SI interactions redistribute the $\nu_e$ among all neutrino and antineutrino flavors, the core is out of weak equilibrium, and the charged current weak interactions will begin to redistribute energy and lepton number. We can use a Boltzmann equation to follow the evolution toward a new beta equilibrium. We do this calculation with a constant density, $\rho_{10}=100$, because collapse timescales are relatively long. $\nu$SI and electromagnetic interactions, occurring on much shorter timescales than the weak interactions, ensure that all species are in \emph{kinetic} equilibrium and can be described with a temperature $T$ and chemical potential $\mu$. For a given species, the rates of change of the number density $dn/dt$ and of the energy density $d\rho/dt$ imply the following evolution of ($T,\mu$)~\cite{EscuderoAbenza:2020cmq}:
\begin{subequations}
\label{eq:dTdmu}
\begin{align}
\label{eq:dT_dt}
\frac{dT}{dt} = \left. \left( \frac{\partial \rho}{\partial \mu} \frac{dn}{dt} - \frac{\partial n}{\partial \mu}\frac{d\rho}{dt} \right) \middle/ \left( \frac{\partial n}{\partial T} \frac{\partial \rho}{\partial \mu} - \frac{\partial n}{\partial \mu}\frac{\partial \rho}{\partial T} \right) \right. \ , \\
\label{eq:dmu_dt}
\frac{d\mu}{dt} = \left. \left(\frac{\partial \rho}{\partial T} \frac{dn}{dt} - \frac{\partial n}{\partial T}\frac{d\rho}{dt} \right) \middle/ \left( \frac{\partial n}{\partial \mu}\frac{\partial \rho}{\partial T} - \frac{\partial n}{\partial T} \frac{\partial \rho}{\partial \mu} \right) \right. \ .
\end{align}
\end{subequations}
We use these to evolve temperatures and chemical potentials for $e^-$, and all neutrino species.
\begin{figure*}[t]
\includegraphics[width=0.68\columnwidth]{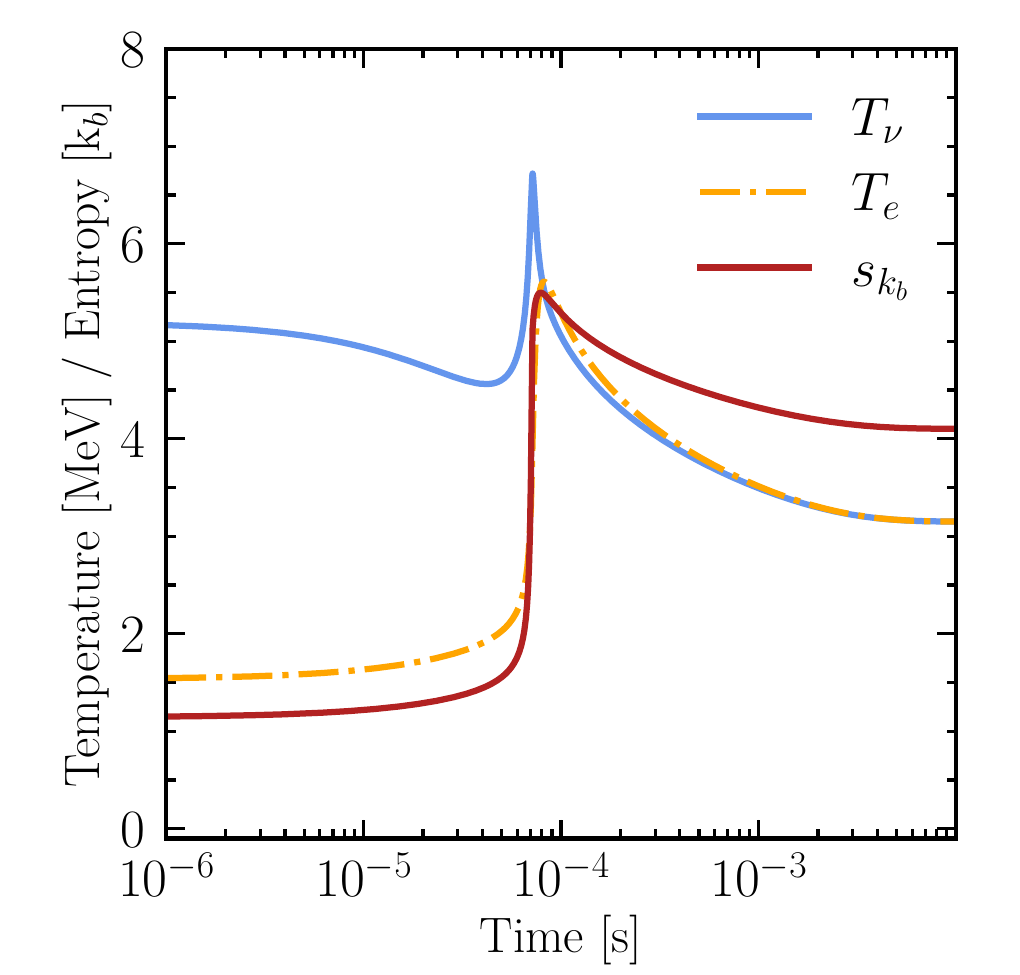} 
\includegraphics[width=0.68\columnwidth]{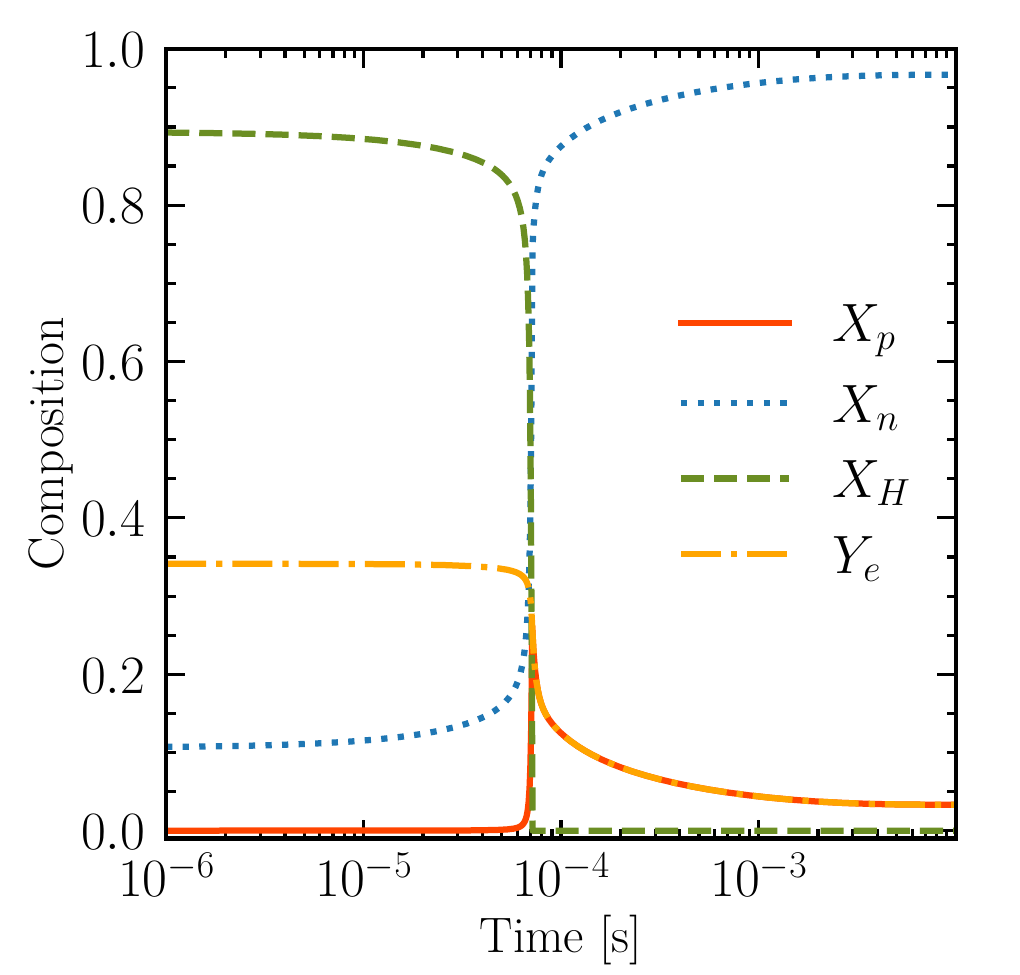} 
\includegraphics[width=0.68\columnwidth]{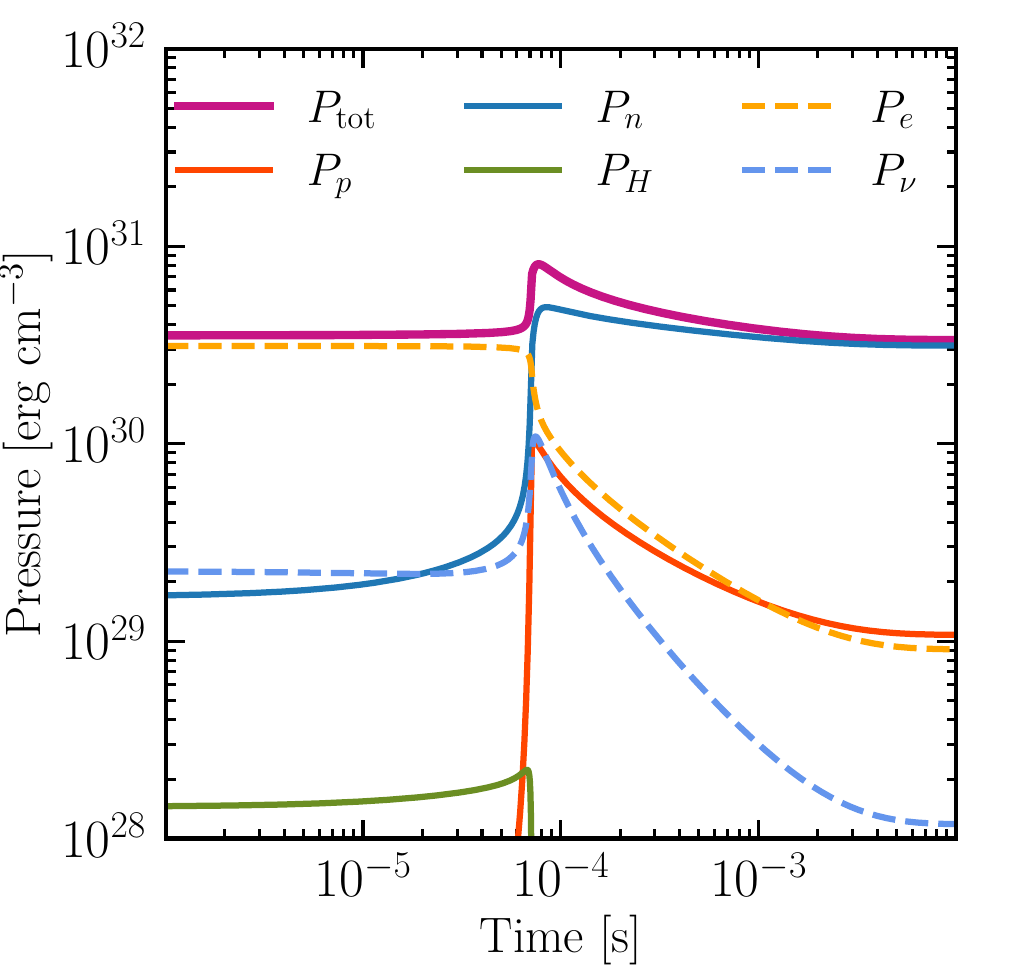} 
\caption{Temporal evolution of the effective temperatures (in MeV) for neutrinos ($T_\nu$), and the matter component ($T_e$), and entropy-per-baryon $s_{k_\mathrm{B}}$ in units of Boltzmann's constant $k_{\rm B}$ (\emph{left panel}); 
the composition of matter: mass fractions for free protons, $X_p$, free neutrons, $X_n$, and heavy nuclei, $X_H$, and electron fraction $Y_e$ (\emph{middle panel}); and the total pressure, $P_{\rm tot}$, and the pressure contributions for free neutrons and protons, $P_{n}$ and $P_{ p}$, respectively, heavy nuclei, $P_{H}$, electrons, $P_e$, and neutrinos of all types, $P_\nu$ (\emph{right panel}). We turn on the LNV $\nu$SI at neutrino trapping, Time zero.} 
\label{Fig1}
\end{figure*} 
When writing Eq.~\eqref{eq:dTdmu} for electrons, $n_{e^-}$ and $\rho_{e^-}$ change because of weak and QED interactions. Since there are initially almost no positrons, the fast number-changing QED interactions $e^- + e^+ \leftrightarrow \gamma + \gamma$ immediately erase any change in $n_{e^+}$ due to weak interactions. In other words, the net change in $e^-$ number density is $d n_{e^-}/dt = d n_{e^-}/dt\rvert_\mathrm{weak} - d n_{e^+}/dt\rvert_\mathrm{weak}$, and the same reasoning can be applied to $\rho_{e^-}$. The evolution equation for $e^-$ quantities is sufficient to describe the whole electromagnetic plasma, since QED interactions ensure $T_{e^-}=T_{e^+}=T_\gamma \equiv T_{e}$ and $\mu_\gamma =0$, $\mu_{e^-}=-\mu_{e^+}\equiv \mu_e$.

Figure~\ref{Fig1} shows the temporal evolution of the temperatures,  matter composition, and pressure. 
The LNV $\nu$SI processes (stronger than the Standard Model weak interaction) equilibrate the $\nu_e, \bar\nu_e, \nu_\mu,\bar\nu_\mu,\nu_\tau,\bar\nu_\tau$ seas, and hence they share a common $T_{\nu}$ and $\mu_\nu$ whose evolution equations are given in the End Matter. Given our large mediator mass, the relevant $\nu$SI are $2 \to 2$ processes (two particles interact to form two particles) which conserve the energy and number densities.

Lowering $\mu_{\nu_e}$ allows enhanced electron capture (see End Matter for the temporal evolution of the weak rates). 
Figure~\ref{Fig1} illustrates the abrupt increase in entropy $s_{k_\mathrm{B}}$ that accompanies neutrino equilibration, and the corresponding abrupt decrease in the heavy nucleus mass fraction $X_H$. 
The baryonic component is always in NSE and the heavy nucleus mass fraction roughly follows the Saha equation,
$X_H \propto{s_{k_\mathrm{B}}^{1-\langle A \rangle} n_p^Z n_n^N \exp(E_{b}/T_e)}$, 
where $n_p$ and $n_n$ are, respectively, the overall proton and neutron number densities, and $E_{b}$ is the binding energy of the nucleus with mass number $\langle A \rangle = N + Z$. As the entropy rises, nuclei with larger $\langle A \rangle$ are disfavored, nearly disappearing when $\Delta s_{k_\mathrm{B}} > $ 3--4.

In the standard picture of core collapse, the electron pressure, and hence $Y_e$, determines the size of the homologous core and the initial shock energy after core bounce~\cite{Brown:1982cpw, Fuller:1981mu}. With LNV $\nu$SI, $Y_e$ decreases significantly. Higher $s_{k_\mathrm{B}}$ and higher matter temperature means a larger free proton mass fraction $X_p$ and that, in turn, favors rapid electron capture with less $\nu_e$ blocking. 
With much higher free neutron mass fraction $X_n$, and lower $Y_e$, the pressure will be dominated by the non-relativistic baryonic component. 

For Cases (2) and (3), when the $\nu$SI enable LNV among $e$ or $e$ and $x$ neutrino flavors, we obtain core physics alterations that are qualitatively similar to those in Case (1), consistent with results found in Refs.~\cite{Kolb:1981mc, Fuller:1988ega}.

We point out that the nuclear and weak interaction physics associated with these high entropy conditions has not been explored. Our liquid-drop-based nuclear physics treatment is likely inadequate (see Supplemental Material).  
However, we can argue that our results are inevitable for strongly LNV $\nu$SI-coupled $\nu_e$ with neutrinos and antineutrinos of multiple flavors that have thermalized with matter and achieved a new beta equilibrium.
The only new equilibrium solution is $\mu_{\nu_e} = 0$. Under the assumption that all nuclei melt, and free nucleons are well described by Boltzmann statistics, the new equilibrium values of $\mu_e$ and $Y_e$, for a common temperature $T_e$, satisfy
\begin{equation}
\label{eq:new_eq}
    \mu_e = \delta m_{np} - T_e \ln{\left(\frac{Y_e}{1-Y_e}\right)} \ .
\end{equation}
This can be solved iteratively, with the result that $Y_e$ will be significantly lower and $X_n$ significantly higher (see Fig.~\ref{Fig:Sol_Analytic} in End Matter). 
Even if the temperature does not increase significantly, the new equilibrium still requires very low $Y_e$ and that likely precludes the existence of a large fraction of heavy nuclei (see Fig.~\ref{Fig8} in Supplemental Material).
We also find that adding $\alpha$ particles to our calculations does not significantly change our results; they are in qualitative agreement with the simulations without the $\alpha$ particles (see Fig.~\ref{Fig:9} in the Supplemental Material).

\prlsection{Discussion and Conclusions}{.}
We have shown that LNV $\nu$SI coupling and mediator mass parameters that are currently unconstrained, but possibly discoverable in future experiments, could drastically alter the CCSN collapse epoch physics. They can do this because they can tap into the huge reservoir of degenerate electron flavor leptons that characterizes the pre-bounce, pre-explosion CCSN.

We point out that a scenario where $2 \rightarrow 4$ processes (two particles interact to form four particles) become important would produce a different evolution in the collapsing core. The lower neutrino energies produced this way could allow neutrinos to escape the in-falling core more readily, speeding up the collapse and decreasing $Y_e$, while leaving the temperature lower. This scenario could lead to a sightly larger heavy nucleus fraction that could persist until a standard nuclear density bounce. The impact of flavor diagonal $\nu$SI $2 \rightarrow 4$ processes has been studied in the context of energy loss arguments for post-bounce evolution in~\cite{Shalgar:2019rqe}.

For LNV $\nu$SI studied in our work, the pressure could be dominated by non-relativistic sources and the temperature will be higher than in the standard case. Is there a thermal bounce~\cite{Fuller:1988ega} before the core reaches nuclear density, or does the collapse proceed to nuclear density, albeit with a larger inner core because of a higher sound speed? Only detailed hydrodynamic and transport simulations with LNV $\nu$SI can answer these questions, but the simulations will have to include the nuclear and weak interaction physics in the exotic high-entropy and low-$Y_e$ conditions found here. Such a simulation could extend the range of LNV $\nu$SI parameters studied here to a region where the separation of the dynamical and the LNV $\nu$SI timescales might be questionable.  These simulations should incorporate detailed neutrino transport coupled to the LNV physics discussed here.

It is clear that there could be significant alterations in both the emergent neutrino and gravitational wave signals for core collapse with this new physics. For example, the electron neutrino burst expected in the standard CCSN case may be replaced by an electron antineutrino burst originating from positron captures on free neutrons. Such a feature was found in a study of a first-order quark-hadron phase transition during the accretion phase of a CCSN~\cite{Sagert:2008ka} and could have consequences for r-process nucleosynthesis~\cite{Fischer:2020xjl}.

This unique new physics provides the exciting prospect of synergy and complementarity between very different experimental probes, from multi-messenger astronomy, including gravitational wave and neutrino detectors, to anticipated accelerator-based laboratory experiments.


\begin{acknowledgments}
\textbf{Acknowledgments.---}
We are grateful for helpful discussions with Frank Deppisch, Miguel Escudero, Bronson Messer, Anthony Mezzacappa, Gail McLaughlin, Evan O'Connor, and Sherwood Richers.
This work was supported in part by National Science Foundation grant PHY-2209578 at UCSD, and by National Science Foundation grant No.\ PHY-2020275: {\it Network for Neutrinos, Nuclear Astrophysics, and Symmetries} (N3AS).  Additional support was provided by the Heising-Simons foundation under grant No.\ 2017-228.
This work was also supported in part by the Department of Energy grant No. DE-AC02-07CHI11359: \emph{Neutrino Theory Network Program}.
This project has received support from the Villum Foundation (Project No. 13164, PI: I. Tamborra), the Danmarks Frie Forskningsfond (Project No. 8049-00038B, PI: I. Tamborra).
Further, the work on this project was supported by the Dutch Research Council (NWO), under project number VI.Veni.222.318, and by Charles University through project PRIMUS/24/SCI/013.
This research was also supported in part by the INT's U.S. Department of Energy grant No. DE-FG02-00ER41132 and by the Center for Theoretical Underground Physics and Related Areas (CETUP), The Institute for Underground Science at Sanford Underground Research Facility (SURF), and the South Dakota Science and Technology Authority.

\end{acknowledgments}

\onecolumngrid
\section{End matter}
\twocolumngrid

\begin{figure}[t]
\includegraphics[width=0.98\columnwidth]{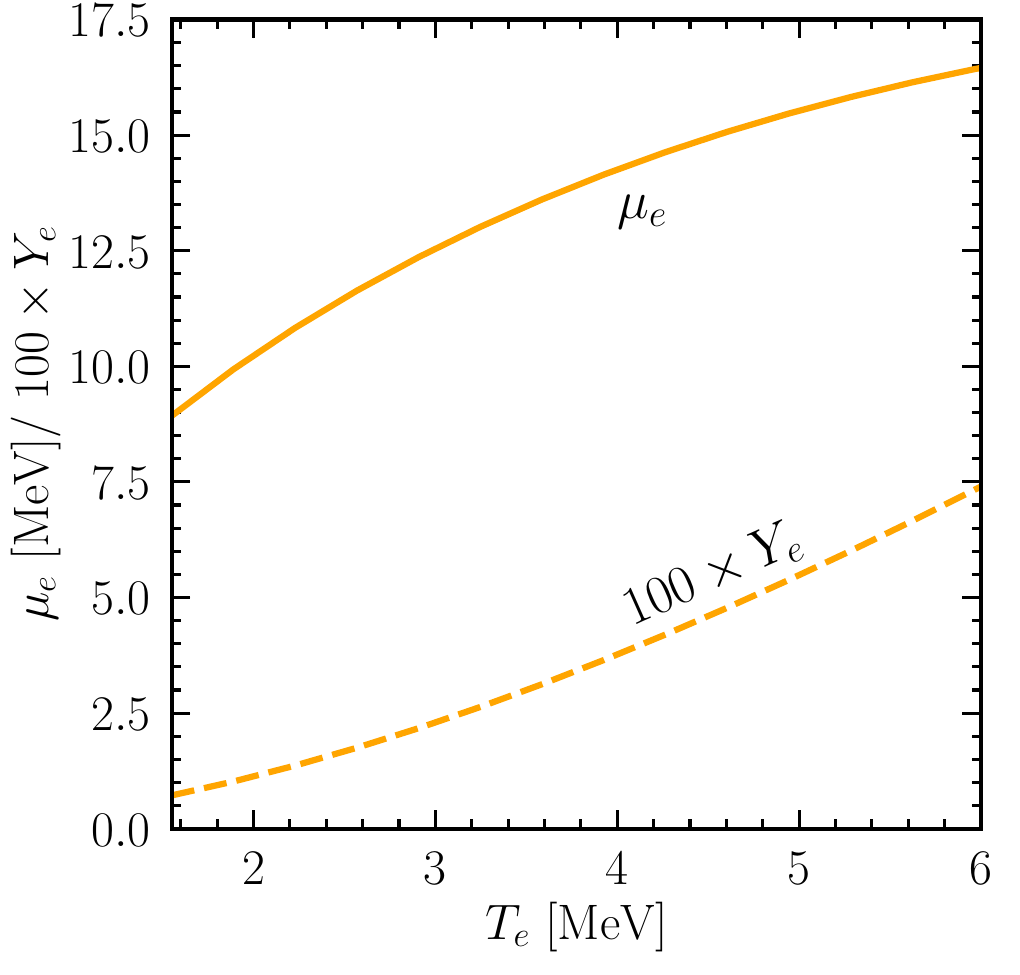} 
\caption{Solutions for the thermal and chemical equilibrium values of $\mu_e$ and $Y_e$ for a given matter temperature $T_e$ and $\mu_{\nu_e} = \mu_{\bar\nu_e} = 0$ in the absence of nuclei.} 
\label{Fig:Sol_Analytic}
\end{figure}

\emph{New Weak Equilibrium with LNV $\nu$SI.---}
Figure~\ref{Fig:Sol_Analytic} shows a solution for a new beta equilibrium with LNV $\nu$SI (new $Y_e$ and $\mu_e$) with negligible chemical potential for neutrinos and no nuclei present. The solution is the root of Eq.~\eqref{eq:new_eq}. Regardless of the final temperature, the new equilibrium has a significantly lower $Y_e$.  

\emph{Weak interaction rates.---}
We review here an approximate way to calculate the charged-current electron, positron, electron neutrinos and antineutrino capture rates on nucleons and nuclei~\cite{Fuller:1980zz, Fuller:1981mt, Fuller:1981mv, Fuller:1981mu, Langanke:2003ii, Juodagalvis:2009pt, Langanke:2020gbk}.

The charged current emission and absorption processes on nucleons and heavy nuclei determine the change in $Y_e$ during the in-fall phase. 
This change can be calculated with $
{dY_e}/{dt} = R_{\nu_e} - R_{\bar\nu_e} - R_{e^-} + R_{e^+}$,
where the rates per baryon of electron, positron, electron neutrino and antineutrino captures on nuclei and nucleons are 
\begin{subequations}
\label{eq:reaction-rates} 
\begin{eqnarray} \label{eq:prob}
R_{\nu_e} &=& \lambda^{n}_{\nu_e}X_n + \lambda^H_{\nu_e} \left( 1 - X_n - X_p\right)/A \ , \\
R_{\bar\nu_e} &=& \lambda^{p}_{\bar\nu_e}X_p + \lambda^H_{\bar\nu_e} \left( 1 - X_n - X_p\right)/A \ , \\
R_{e^-} &=& \lambda^{p}_{e^-}X_p + \lambda^H_{e^-} \left( 1 - X_n - X_p\right)/A  \ , \\
R_{e^+} &=& \lambda^{n}_{e^+}X_n + \lambda^H_{e^+} \left( 1 - X_n - X_p\right)/A \ ,
\end{eqnarray}
\end{subequations}
with $X_n$, $X_p$ and $X_H = 1 - X_n - X_p$ the fractions of free neutrons, free protons, and heavy nuclei in the absence of alpha particles, and 
\begin{equation}
    \lambda_{i}^{j} = \int \sigma_{ij} \frac{dn_i}{dE_i} dE_i  = \log 2 \frac{f_{ij}}{ft_{ij}}\ ,
\end{equation}
where $ \sigma_{ij}$ is the cross section for lepton capture between the parent state $i$ and daughter state $j$,
$ft_{ij}$ is the comparative half-life, and the differential particle number density of the relativistic particles $l$ for homogenous and isotropic distributions is given by
${dn_l}/{dE_l} \approx {g_l}/({2\pi^2}) {E_l^2} S_l(E_l)  \ ,$
where $g_l$ is the statistical weight and $S_l$ is the distribution of particles $l$ over the energy --- for fermions in thermal and chemical equilibrium, i.e, a Fermi-Dirac distribution. The phase space factor $f_{ij}$ for the electron and neutrino captures on the free nucleons and heavy nuclei can be calculated following~Ref.~\cite{Fuller:1980zz, Fuller:1981mt, Fuller:1981mv, Fuller:1981mu, Fuller:1995ih} as
\begin{equation}
\label{eq:phase-space-e}
    f_{e}^{p,H} = \frac{1}{m_e^5} \int_Q^\infty G E_e^2 (E_e - Q)^2 S_e (1-S_\nu) dE_e \ ,
\end{equation}
and
\begin{equation}
\label{eq:phas-space-factor}
f_{\nu_e}^{n, H} = \frac{1}{m_e^5}\int_{E_\mathrm{th}}^\infty G E_\nu^2 (Q + E_\nu)^2 S_\nu (1- S_e) dE_\nu \ ,
\end{equation}
where $S_e$ and $S_\nu$ are the Fermi-Dirac functions for electrons and neutrinos, and $G$ is the Coulomb barrier penetration factor as defined in the Eq.~9 of Ref.~\cite{Fuller:1981mu}.
The resonance nuclear $Q$-value can be approximated with $Q = M_Y - M_X + E_\mathrm{ex}$~\cite{Fuller:1981mu},
where $M_X$ ($M_Y$) is the mass of the parent (daughter) nucleus and $E_\mathrm{ex}$ is the excitation energy for the Gamow-Teller transition in the daughter nucleus. We use the approximation given in~Ref.~\cite{Fuller:1981mv} to calculate the contribution only from the leading transitions such that $Q \approx \hat\mu + \delta m_{np} + E_\mathrm{ex}$, where $\hat\mu$ is the difference between the kinetic chemical potentials of neutrons and protons, but setting $E_\mathrm{ex} = 0$. For $e^+$ and $\bar\nu_e$ captures $Q \rightarrow -Q$, the Fermi function and integration thresholds in Eqs.~\eqref{eq:phase-space-e} and~\eqref{eq:phas-space-factor} must be adjusted.

\begin{figure*}[t]\includegraphics[width=0.68\columnwidth]{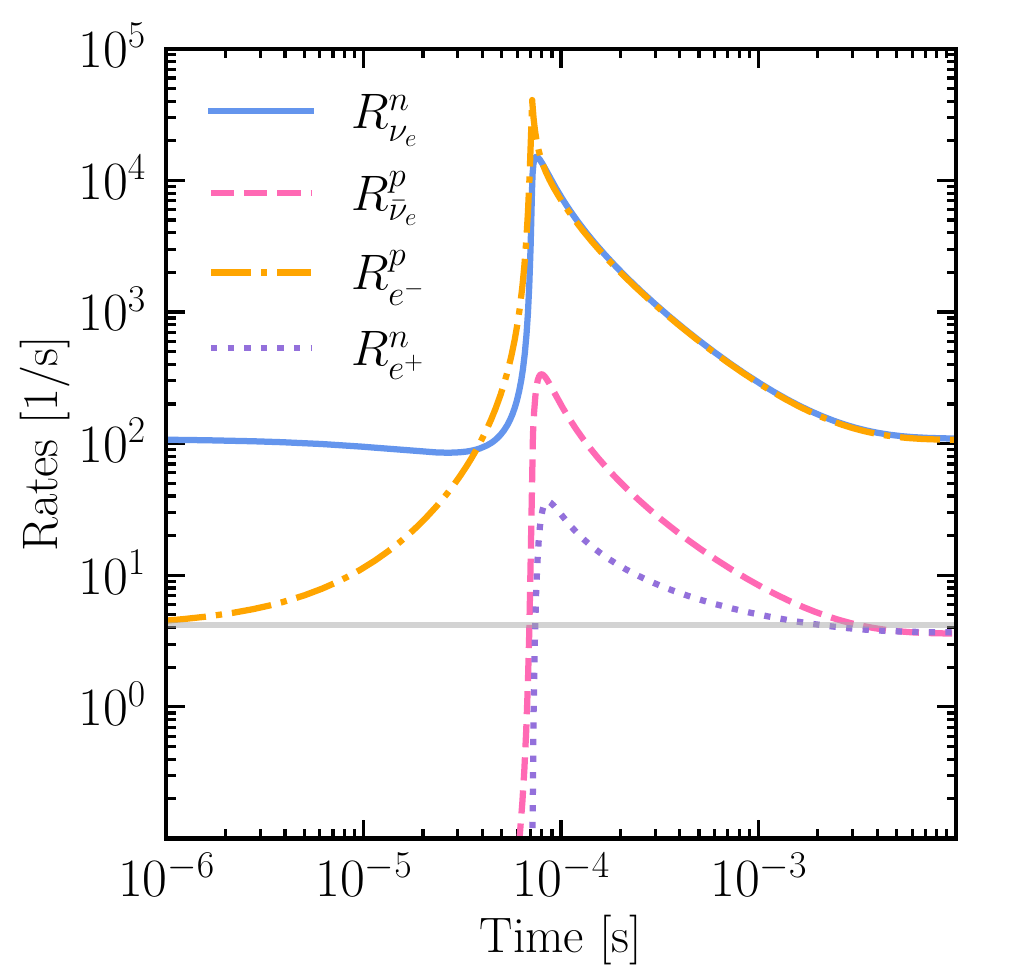} 
\includegraphics[width=0.68\columnwidth]{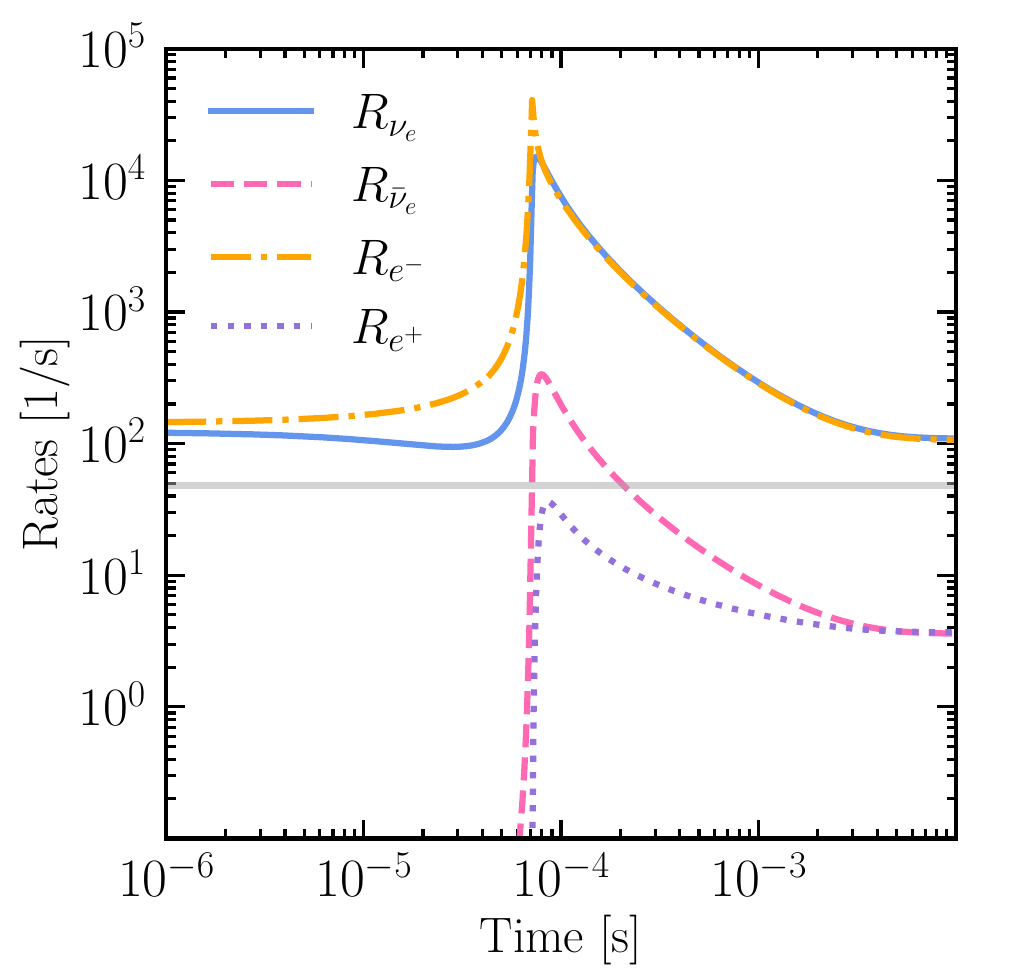} 
\includegraphics[width=0.68\columnwidth]{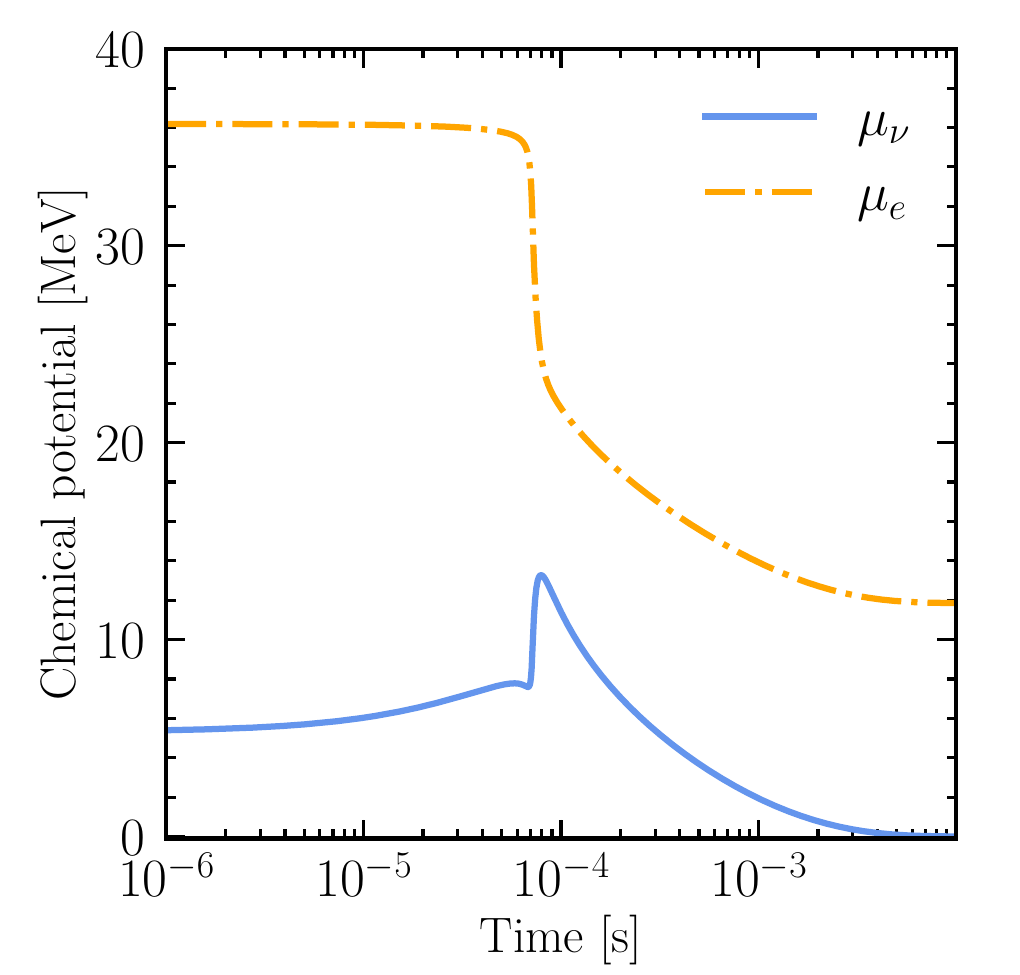} 
\caption{Temporal evolution of the charged-current weak capture rates of electrons, positrons, electron neutrinos, and electron antineutrinos, and chemical potentials for Case (1). \emph{Left panel:} Captures on free nucleons; \emph{Middle panel:} Total capture rates; \emph{Middle panel:} $e^-$, $\nu$ chemical potentials. Horizontal lines show the electron neutrino and electron capture rates in the SM case. The asymptotic new steady state solution is reached within $\mathcal{O}(1)$$\;$ms.}
\label{Fig:Weak_Rates}
\end{figure*}

Figure~\ref{Fig:Weak_Rates} shows the temporal evolution of the charged-current capture rates altered by the presence of the LNV $\nu$SI. The left panel corresponds to the capture rates on free nucleons, and the right panel corresponds to the total capture rates, i.e., including the captures on free nucleons and nuclei. 
The increased temperature of the $\nu_e$ sea makes $\nu_e$ captures on free neutrons and nuclei more efficient, which heats the electrons.  
Also, the smaller $\mu_{\nu_e}$ opens up phase space for electron capture on nuclei, increasing the rate for this process. In contrast, $e^-$ capture on free protons is insignificant at the beginning because their small mass fraction is the limiting factor in the overall capture rate. 
In our simulation, a sufficient amount of entropy to melt the nuclei is generated while attaining the new beta equilibrium; the remaining evolution is dominated by electron captures on free protons.

We have ignored neutrino captures $\nu_e + A(Z-1, N+1) \rightarrow A^*(Z, N) + e^-$  to the highly excited Gamow-Teller resonance state and isobaric analogue state in the daughter nucleus $A^*(Z, N)$~\cite{Fuller:1981mv}. Though the Gamow-Teller and Fermi strength in these channels is very large, the rate for these channels will be threshold-suppressed and will produce relatively low-energy final state electrons that will tend to be blocked. Moreover, addition of this sub-leading channel would serve only to increase entropy generation, augmenting the evolution towards melting heavy nuclei.

\emph{LNV $\nu$SI without Scattering Kernel.---}
Given the fast equilibration by strong LNV $\nu$SI, the ensemble of (anti)neutrinos is a tightly-coupled fluid for which we can rewrite Eqs.~\eqref{eq:dT_dt} -- \eqref{eq:dmu_dt} for a common temperature $T_\nu$ and a common chemical potential $\mu_\nu$. Indeed, the Boltzmann equations read~\cite{EscuderoAbenza:2020cmq}:
\begin{subequations}
\label{eq:coupled_neutrinos}
    \begin{align}
        \sum_{\alpha}{\left(\frac{dn_{\nu_\alpha}}{dt} + \frac{d n_{\bar{\nu}_\alpha}}{dt}\right)} &= \frac{\delta n_{\nu}}{\delta t} \, , \\
        \sum_{\alpha}{\left(\frac{d \rho_{\nu_\alpha}}{dt} + \frac{d \rho_{\bar{\nu}_\alpha}}{dt}\right)} &= \frac{\delta \rho_{\nu}}{\delta t} \, ,       
    \end{align}
\end{subequations}
where $\alpha$ is one of the $N_F$ neutrino flavors and the right-hand sides are the total variation of (anti)neutrino number and energy densities through the charged-current weak interactions. Writing $n_\nu$ and $\rho_\nu$ the number and energy density of each neutrino species, identical for all species because of $\nu$SI, we can apply the chain rule similarly to Eqs.~\eqref{eq:dT_dt}--\eqref{eq:dmu_dt}
for all species. The evolution equations for the temperature and chemical potential of the (anti)neutrino fluid then read
\begin{subequations}
\label{eq:dTdmunu}
\begin{align}
    \frac{d T_\nu}{dt} &= \frac{\dfrac{\partial \rho_\nu}{\partial \mu_\nu} \dfrac{\delta n_{\nu}}{\delta t} - \dfrac{\partial n_\nu}{\partial \mu_\nu} \dfrac{\delta \rho_{\nu}}{\delta t}}{2 N_F\left(\dfrac{\partial n_\nu}{\partial T_\nu} \dfrac{\partial \rho_\nu}{\partial \mu_\nu} - \dfrac{\partial n_\nu}{\partial \mu_\nu}\dfrac{\partial \rho_\nu}{\partial T_\nu}\right)} \label{eq:1nu} \, , \\
    \frac{d \mu_\nu}{dt} &= \frac{\dfrac{\partial \rho_\nu}{\partial T_\nu} \dfrac{\delta n_{\nu}}{\delta t} - \dfrac{\partial n_\nu}{\partial T_\nu} \dfrac{\delta \rho_{\nu}}{\delta t}}{2 N_F\left(\dfrac{\partial n_\nu}{\partial \mu_\nu} \dfrac{\partial \rho_\nu}{\partial T_\nu} - \dfrac{\partial n_\nu}{\partial T_\nu} \dfrac{\partial \rho_\nu}{\partial \mu_\nu}\right)} \label{eq:2nu} \, .      
\end{align}
\end{subequations}

\onecolumngrid
\phantom{i}
\twocolumngrid
\phantom{i}
\bibliography{INFALL-BSM}
\phantom{i}
\clearpage

\appendix


\clearpage
\newpage
\onecolumngrid

\centerline{\large {Supplemental Material for}}
\medskip

{\centerline{\large \bf{Lepton Number Violation in the Neutrino Sector}}}
{\centerline{\large \bf{Could Change the Prospects for Core Collapse Supernova Explosions}}}
\medskip
{\centerline{Anna~M.~Suliga, Patrick Chi-Kit Cheong,  Julien Froustey, George~M.~Fuller, {Luk\'{a}\v{s} Gr\'{a}f}, Kyle Kehrer,}}
{\centerline{ Oliver Scholer,  and Shashank Shalgar}}
\bigskip
\bigskip

Here, we provide information that is not necessary to understand the primary message of our work, but that may promote further developments and provide a deeper explanation of some of the points made in the main text.  

\section{Lepton Number and Lepton Flavor Violating $\nu$SI for Two and Single Flavor in CCSN}

\begin{figure}[!ht]
\includegraphics[width=0.32\columnwidth]{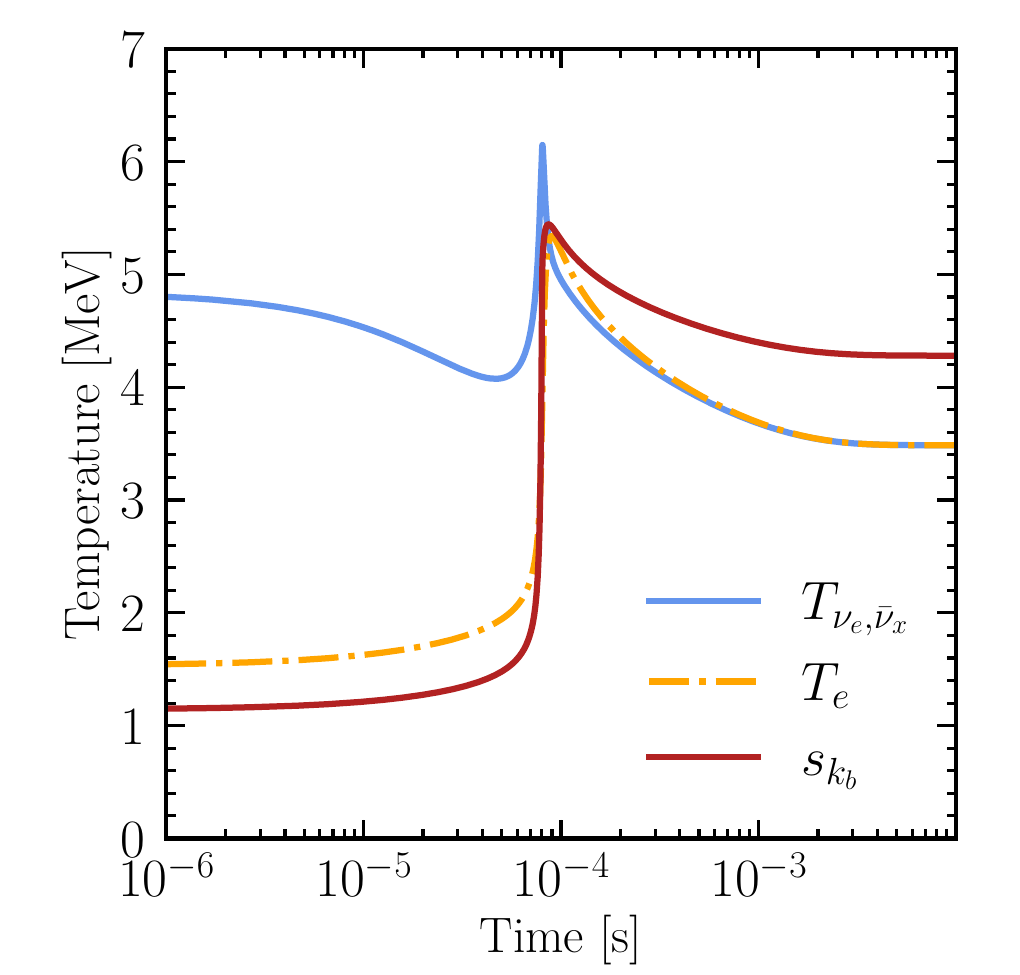} 
\includegraphics[width=0.32\columnwidth]{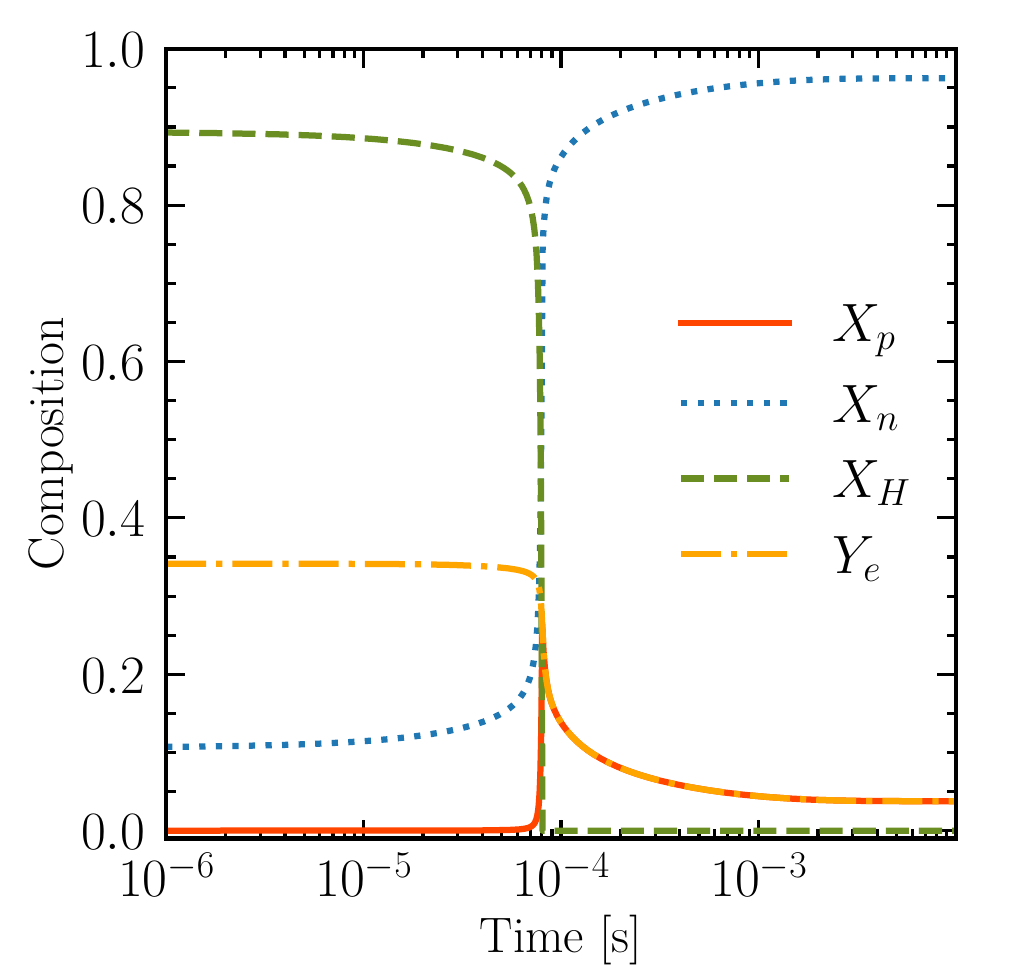} 
\includegraphics[width=0.32\columnwidth]{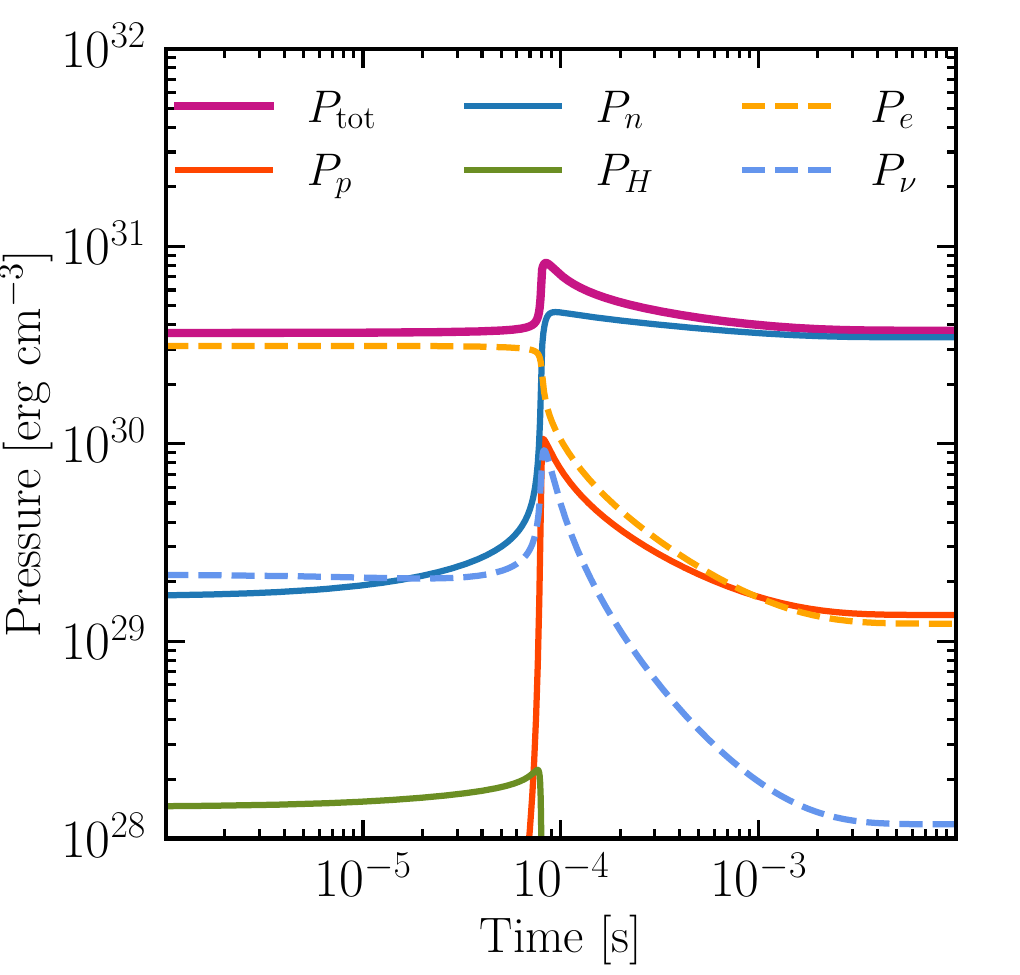} 
\includegraphics[width=0.32\columnwidth]{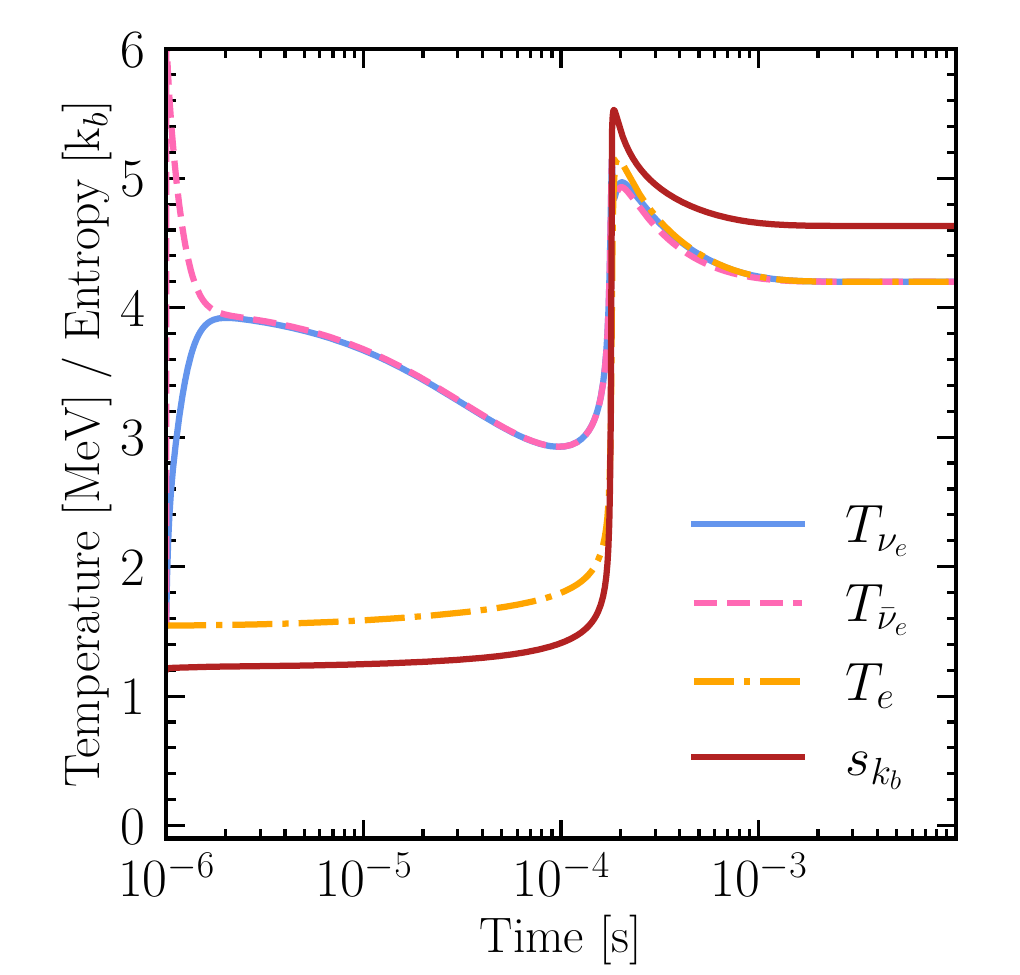} 
\includegraphics[width=0.32\columnwidth]{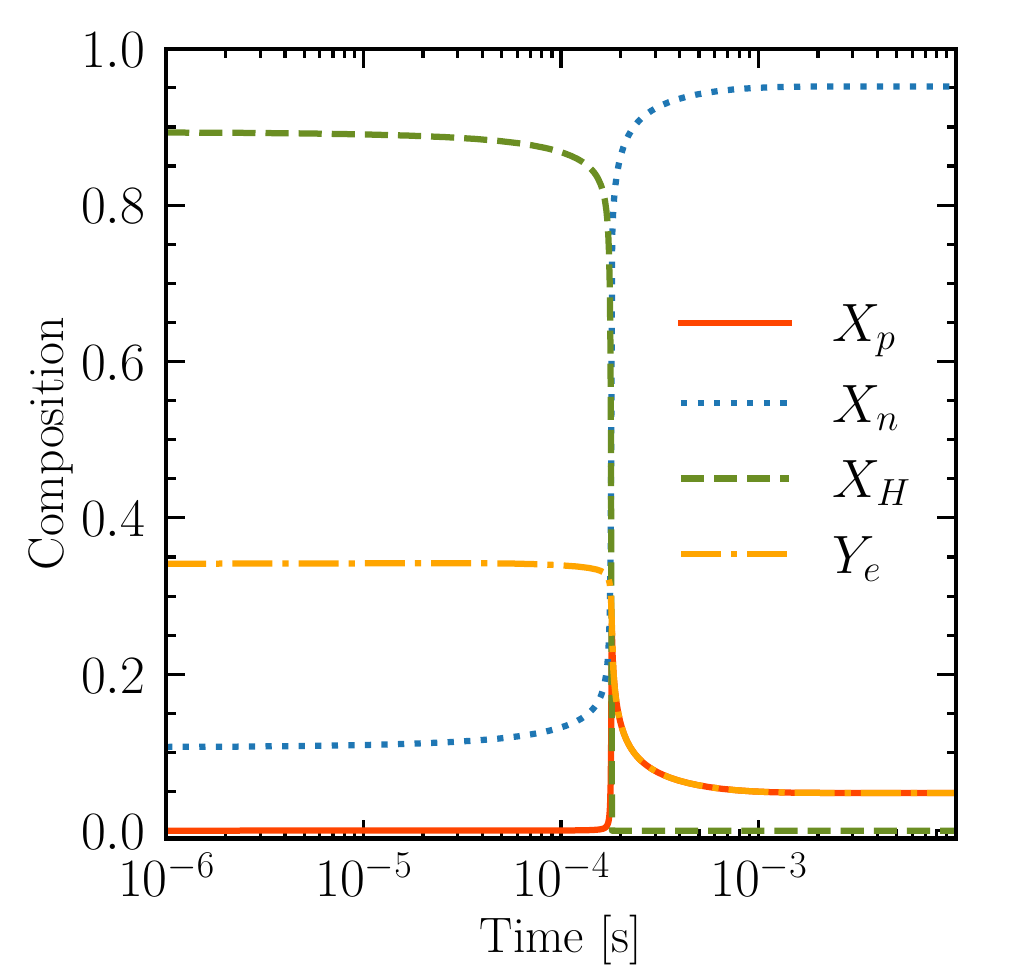} 
\includegraphics[width=0.32\columnwidth]{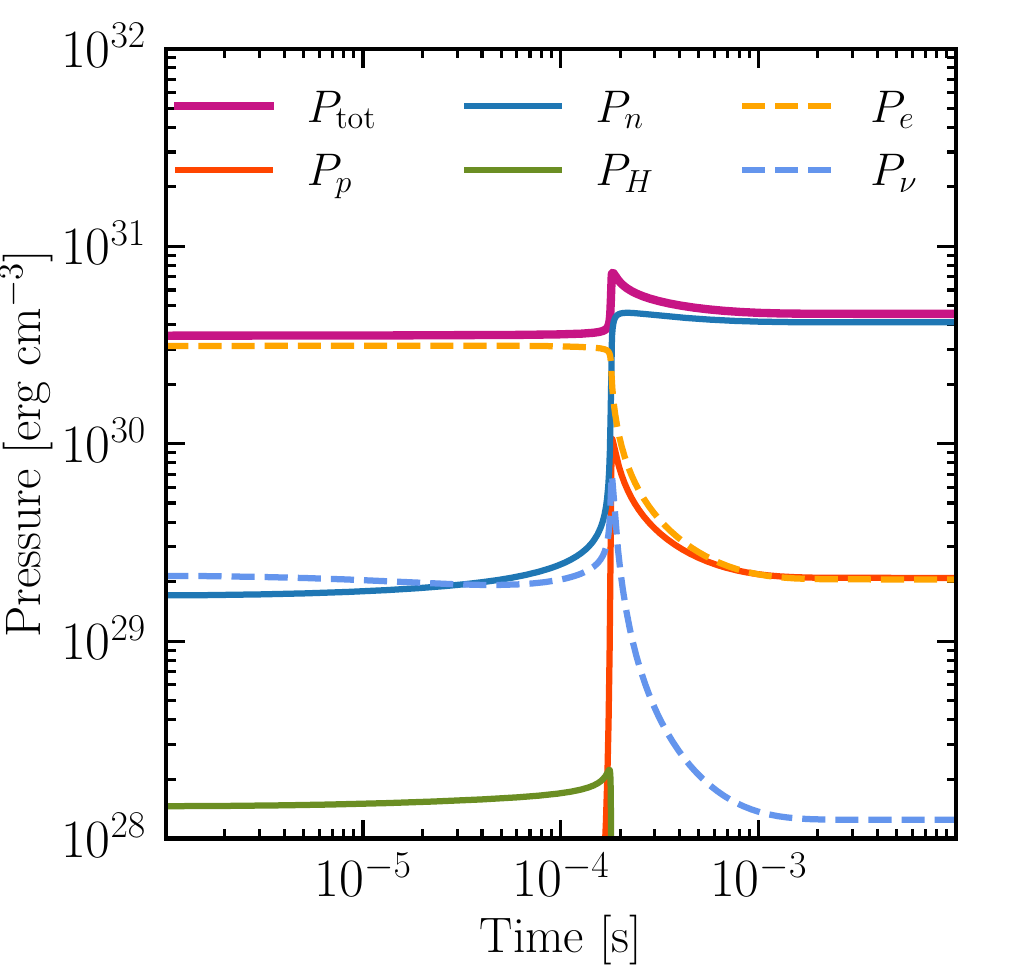} 
\caption{ Temporal evolution of the effective temperatures (in MeV) for neutrinos ($T_{\nu_i}$), and the matter component ($T_e$), and entropy-per-baryon $s_{k_\mathrm{B}}$ in units of Boltzmann's constant $k_{\rm B}$ (\emph{left panel}); the composition of matter: mass fractions for free protons, $X_p$, free neutrons, $X_n$, and heavy nuclei, $X_H$, and electron fraction $Y_e$ (\emph{middle panel}); and the total pressure, $P_{\rm tot}$, and the pressure contributions for free neutrons and protons, $P_{n}$ and $P_{ p}$, respectively, heavy nuclei, $P_{H}$, electrons, $P_e$, and neutrinos of all types, $P_\nu$ (\emph{right panel}). \emph{Top panels} showcase Case (2) and \emph{bottom panels} Case (3). We turn on the LNV $\nu$SI at neutrino trapping, Time zero.}
\label{Fig5}
\end{figure}

The top panels of Fig.~\ref{Fig5} show the temporal evolution of the neutrino and electron temperatures together with the entropy-per-baryon (left panel), the temporal evolution of the matter composition (middle panel), and the pressure (right panel) for the LNV Lepton Flavor Violating (LFV) $\nu$SI with the same coupling for four neutrino species, referred to in the main text as Case (2), whereas the bottom panels illustrate the same quantities for LNV $\nu$SI in Case (3), where only $\nu_e$ and $\bar{\nu}_e$ equilibrate by LNV $\nu$SI.

We use the approximation for strong LNV $\nu$SI described in the \emph{End Matter} to evolve only a single temperature $T_{\nu_e,\nu_x}$ and chemical potential $\mu_{\nu_e,\nu_x}$ for all four neutrinos and antineutrinos in Case (2). The final temperature of the electrons and neutrino fluid is slightly larger than in Case (1), where all of the six (anti)neutrino species participate in LNV $\nu$SI.

For Case (3), we included the LNV $\nu$SI kernels in the evolution equations for $\nu_e$ and $\bar\nu_e$ temperatures and chemical potentials [Eqs.~\eqref{eq:dTdmu}]. The strong LNV $\nu$SI processes we consider here equilibrate the $\nu_e$ and $\bar\nu_e$ seas on a very short timescale, and hence  $T_{\nu_e}$ and $T_{\bar\nu_e}$, as well as $\mu_{\nu_e}$ and $\mu_{\bar\nu_e}$. This justifies using a single temperature and chemical potential to evolve those quantities for the strongly coupled by the LNV $\nu$SI neutrino species.

In addition, the matter and neutrinos attain the new weak equilibrium faster in Case (3) than in Cases (1) and (2), because in Case (3) all of the equilibrated by LNV $\nu$SI species of neutrinos can communicate with the matter through charged-current weak interactions.

\begin{figure}[t]
\includegraphics[width=0.32\columnwidth]{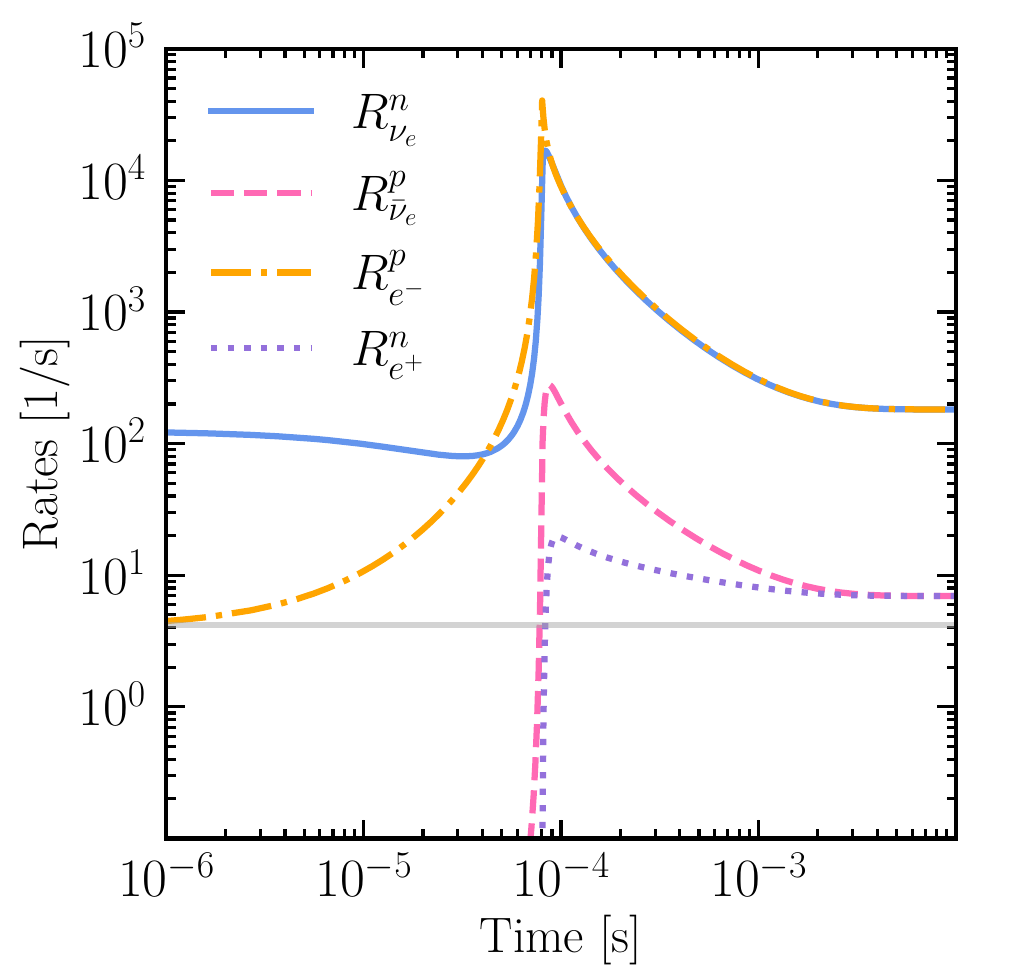} 
\includegraphics[width=0.32\columnwidth]{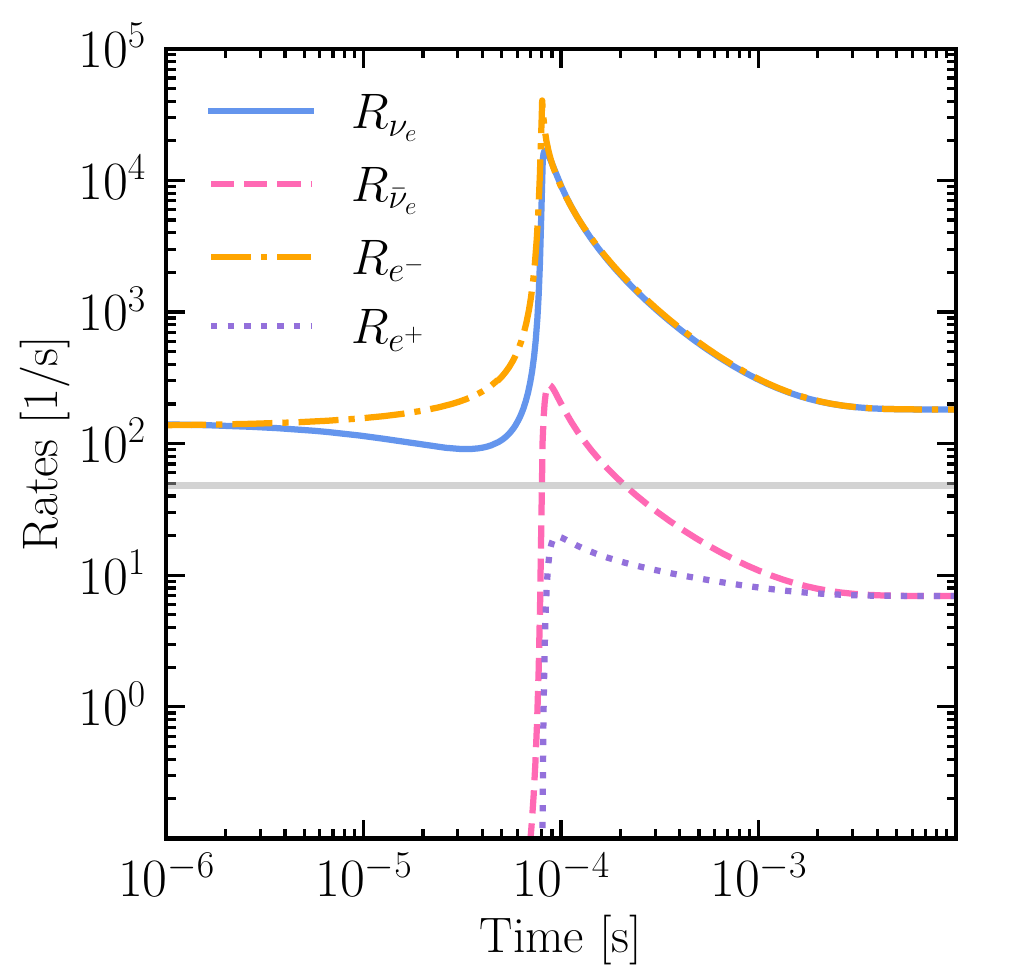} 
\includegraphics[width=0.32\columnwidth]{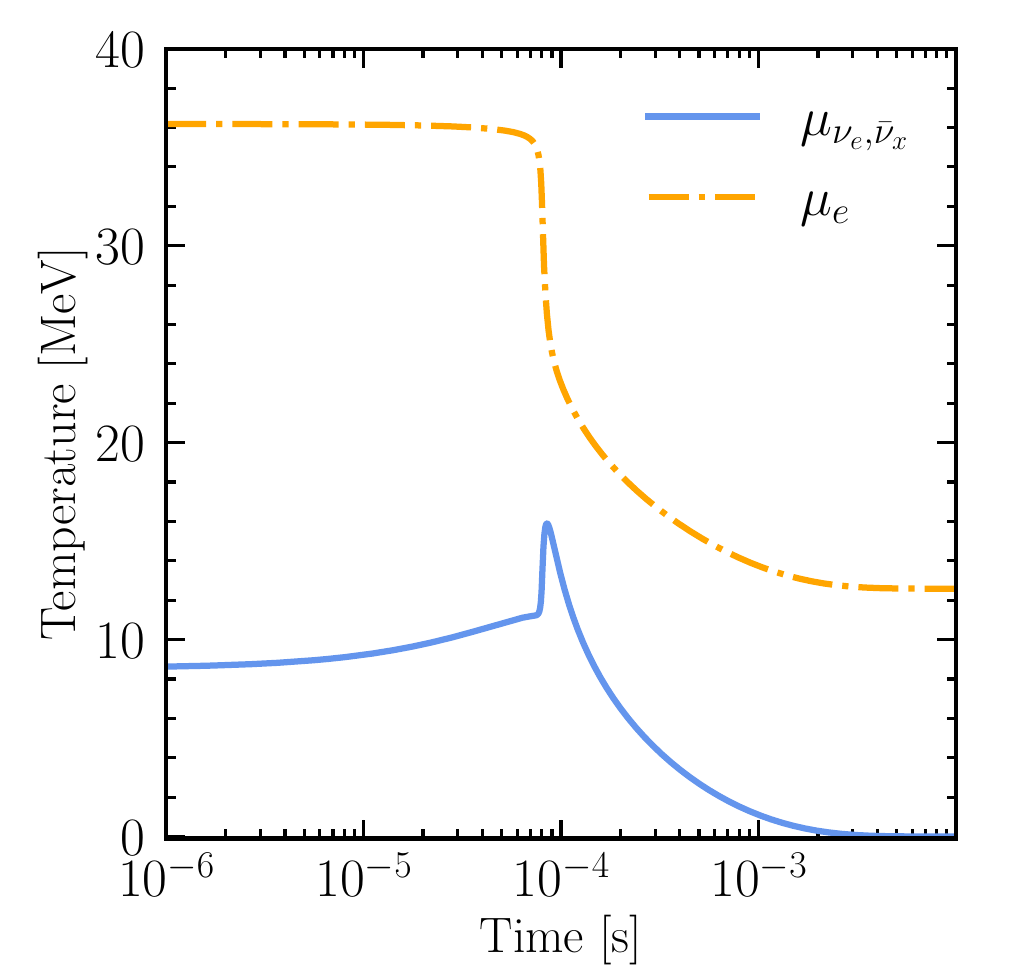} 
\includegraphics[width=0.32\columnwidth]{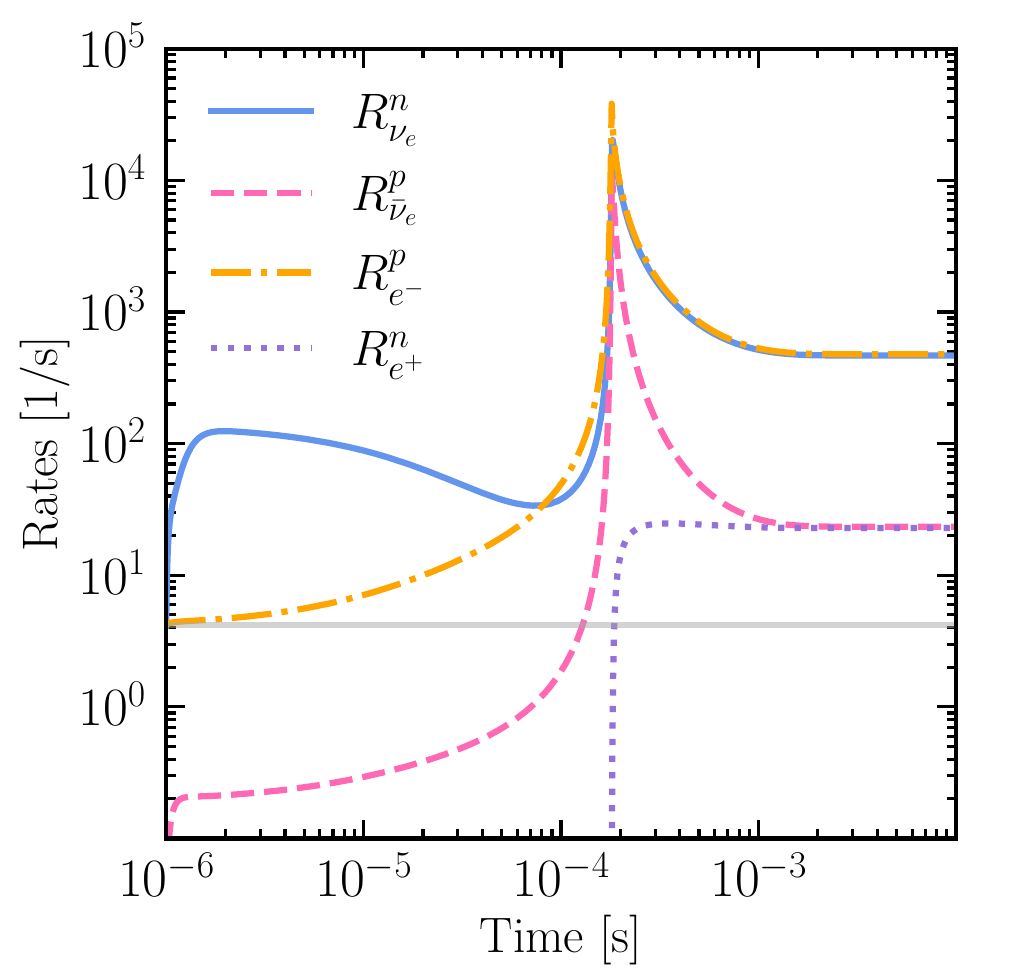} 
\includegraphics[width=0.32\columnwidth]{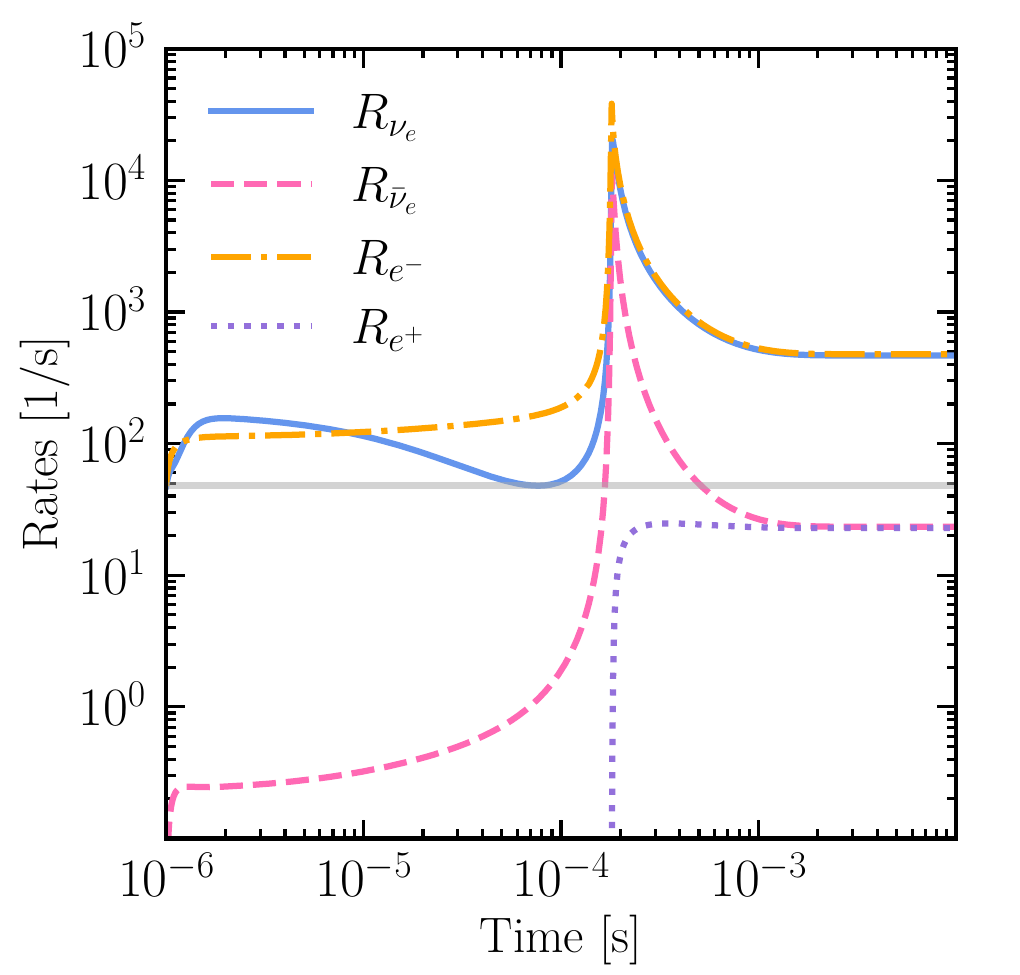} 
\includegraphics[width=0.32\columnwidth]{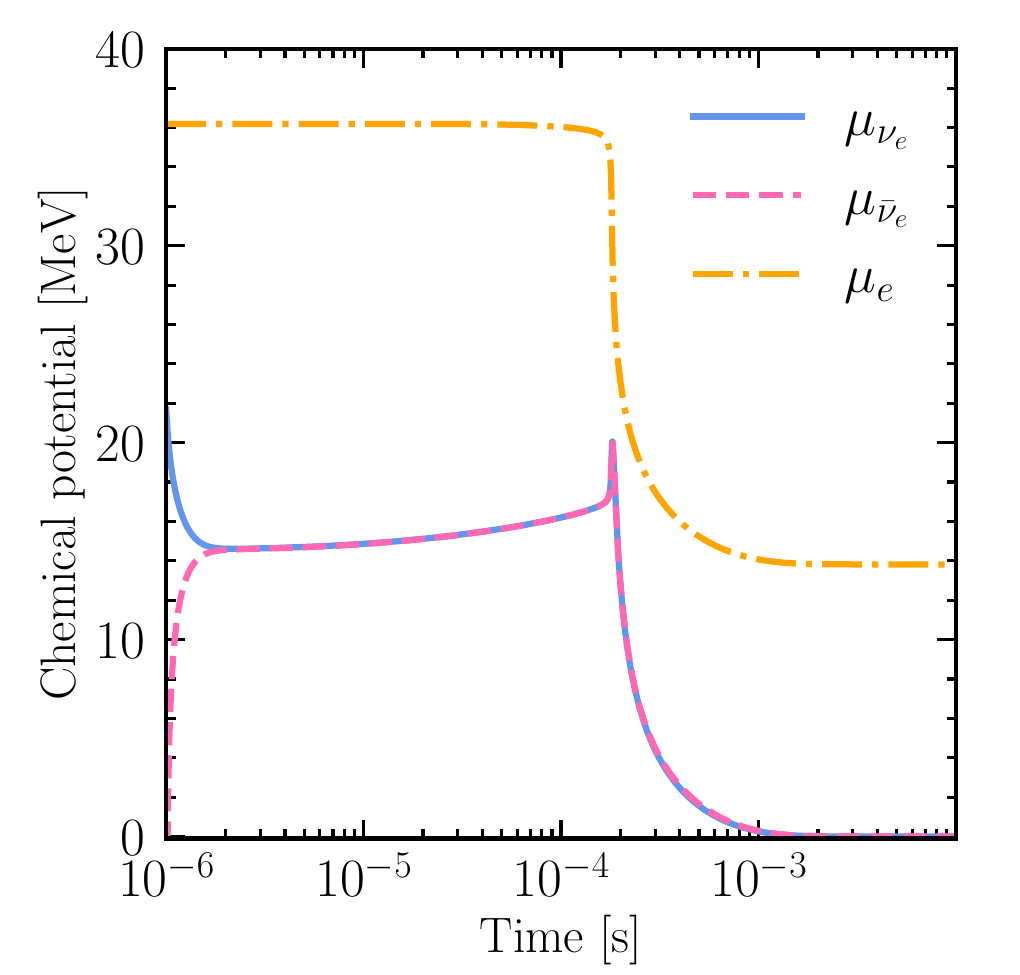} 
\caption{Temporal evolution of the charged-current weak capture rates of electrons, positrons, electron neutrinos, and electron antineutrinos, and chemical potentials for LNV $\nu$SI between two flavors $\nu_e \rightleftharpoons \nu_e \bar\nu_x$, i.e, Case (2) [\emph{top panels}] and for LNV $\nu$SI in $\nu_e$ only, i.e., Case (3) [\emph{bottom panels}]. \emph{Left panel:} Captures on free nucleons; \emph{Middle panel:} Total capture rates; \emph{Middle panel:} $e^-$ and neutrino chemical potentials. Horizontal lines show the electron neutrino and electron capture rates in the SM case. The asymptotic new steady state solution is reached within $\mathcal{O}(1)$$\;$ms.}
\label{Fig:7}
\end{figure} 

\section{Partial derivatives of the Fermi Dirac integrals}
The evolution equations~\eqref{eq:dTdmu} and \eqref{eq:dTdmunu} require the use of the partial derivatives of the number and energy densities for particles following Fermi-Dirac distributions. These are given by~\cite{EscuderoAbenza:2020cmq}
\begin{subequations}
\label{eq:FD_partial_deriv}
\begin{align}
\frac{\partial n}{\partial T}  &=  \frac{g}{2\pi^2}  \int dE E \sqrt{E^2 - m^2}\frac{(E - \mu ) }{4T^2} \cosh^{-2}\left(\frac{E - \mu }{2 T}\right)  , \\
\frac{\partial \rho}{\partial T}  &=  \frac{g}{2\pi^2}  \int   dE  E^2 \sqrt{E^2 - m^2}  \frac{(E - \mu ) }{4 T^2} \cosh^{-2}\left(\frac{E - \mu }{2 T}\right) , \\
\frac{\partial n}{\partial \mu}   &=  \frac{g}{2\pi^2}  \int   dE  E \sqrt{E^2 - m^2} \left[{2 T \cosh \left(\frac{E - \mu }{T}\right) + 2 T}\right]^{ - 1}  , \\
\frac{\partial \rho}{\partial \mu}  &=  \frac{g}{2\pi^2}  \int   dE  E^2 \sqrt{E^2 - m^2}  \left[{2 T \cosh \left(\frac{E - \mu }{T}\right) + 2 T}\right]^{ - 1}   ,
\end{align}
\end{subequations}
where the integral limits are $(m, \infty)$, $m$ is the mass of the fermion, $g$ is the statistical weight, $\mu$ is the total chemical potential, $T$ is the temperature, and $E$ is the energy.

\section{Approximate Treatment of the Equation of State}
To gather the extent of the effects of altered electron capture rates during the in-fall of the supernova core, we utilize the equation of state (EOS) prescription from Refs.~\cite{Bethe:1979zd, Fuller:1981mu}. The equations for the neutron kinetic chemical potential $\mu_n$, mean nuclear mass $\langle A \rangle$, difference between the neutron and proton kinetic potentials $\hat\mu$, and nuclear surface energy are  
\begin{subequations}
\label{eq:equations-nuclear-properties} 
\begin{align}\label{eq:1}
\mu_n &= -16 + 125(0.5 - \tilde Y_e) - 150(0.5 - \tilde Y_e)^2 - 2W_\mathrm{surf}\langle A \rangle^{-1/3} \frac{1 - 2 \tilde Y_e}{1- \tilde Y_e}, \\ \label{eq:mu_n}
\langle A \rangle &= \frac{194(1 - \tilde Y_e)^2}{1 - 0.236 \rho^{1/3}} \ , \\
\label{eq:3}
\hat{\mu} &= 250(0.5 - \tilde Y_e) - \frac{W_\mathrm{surf}}{\langle A \rangle^{1/3}} \left(\frac{1}{Y_e} + {2}{\tilde Y_e} \frac{1-2 \tilde Y_e}{1- \tilde Y_e} \right)\ , \\
\label{eq:4}
W_\mathrm{surf} &= 290 \tilde Y_e^2 (1 - \tilde Y_e)^2 \ .
\end{align}
\end{subequations}
At low fraction of free protons, the effective electron fraction $\tilde Y_e \equiv Z/A \approx Y_e/(1-X_n)$, however, as the $X_p$ starts to increase, one has to use the full expression $\tilde Y_e \equiv Z/A = (Y_e - Y_p)/(1-X_n-Y_p)$. Once the temperature rises significantly and the electron fraction decreases our EOS prescription may no longer be valid. However, we have checked that the Lattimer and Swesty EOS~\cite{Lattimer:1991nc} also restricts the existence of nuclei to approximately 4~MeV for density $\rho_{10}=100$ (see also Fig.~\ref{Fig:9}).

\section{$\nu$SI Scattering Kernel}
\label{sec:Kernel}
To test the validity of our assumption of treating the several species of neutrinos as a single sea characterized by a single temperature and chemical potential, we modify the evolution equations Eqs.~\eqref{eq:dTdmu} used in Case (3) for both $\nu_e$ and $\bar\nu_e$ to include the collision terms for the lepton number violating processes $\nu_e + \nu_e \rightleftarrows \bar{\nu}_e + \bar{\nu}_e$.
In general the collision term for two particle scattering inelastically off each other $1+2 \rightarrow 3 + 4$ is expressed with~\cite{Hannestad:1995rs, Dolgov:1997mb, Grohs:2015tfy, Froustey:2022sla} 
\begin{equation}
\label{eq:Collisional_integral}
I_{coll} = \frac{1}{2E_1}\sum \int \frac{d^3 p_2}{2E_2 (2\pi)^3}
\frac{d^3 p_3}{2E_3 (2\pi)^3} \frac{d^3 p_4}{2E_4 (2\pi)^3}
(2\pi)^4\delta^{(4)} (p_1+p_2-p_3-p_4) F(S_1,S_2,S_3,S_4)
\mathcal{S}\, \langle|\mathcal{M}|^2\rangle \ ,
\end{equation}
where $\langle|\mathcal{M}|^2\rangle$ is the $\nu$SI LNV interaction matrix element squared and summed over spins of
particles 2, 3, 4, the blocking factor is $F = S_3 S_4 (1-S_1)(1-S_2)-S_1 S_2 (1-S_3)(1-S_4)$, and $\mathcal{S}$
is the symmetrization factor taking care of pairs of identical 
particles.
We have calculated the neutrino-neutrino scattering kernels using the reductions of the nine dimensional integrals to the two dimensional with the use of the integral representation of the delta function~\cite{Dolgov:1997mb}.

Under the assumption that the CC reactions and electron scattering reactions have timescales much longer than the $\nu$SI we can solve the system of two times number of energy modes differential equations which couple neutrino and antineutrino distributions. The steady state solution for the neutrino and antineutrino distribution is again a Fermi-Dirac distribution that conserves the number of particles and their total energy. The new temperature and chemical potential of neutrinos and antineutrinos, hence, can also be found by solving the energy conservation and number conservation equations 
\begin{subequations}
\label{eq:conservation-T-mu} 
\begin{align}
\sum_{\alpha} \int dE E^2 \left(S_{\nu_\alpha} \left(T_e, \mu_{\nu_\alpha}\right)  + S_{\bar\nu_\alpha} \left(T_e, \mu_{\bar\nu_\alpha}\right) \right)  &= 2N_F \int dE E^2 S_{\nu_e} (T_\nu, \mu_{\nu}) \ , \\
\sum_{\alpha} \int dE E^3 \left(S_{\nu_\alpha} \left(T_e, \mu_{\nu_\alpha}\right) + S_{\bar\nu_\alpha} \left(T_e, \mu_{\bar\nu_\alpha} \right)\right) &= 2N_F \int dE E^3 S_{\nu_e} (T_\nu, \mu_{\nu}) \ .
\end{align}
\end{subequations}
%

\section{New Weak Equilibrium with LNV $\nu$SI}
\begin{figure}[!ht]
\includegraphics[width=0.45\columnwidth]{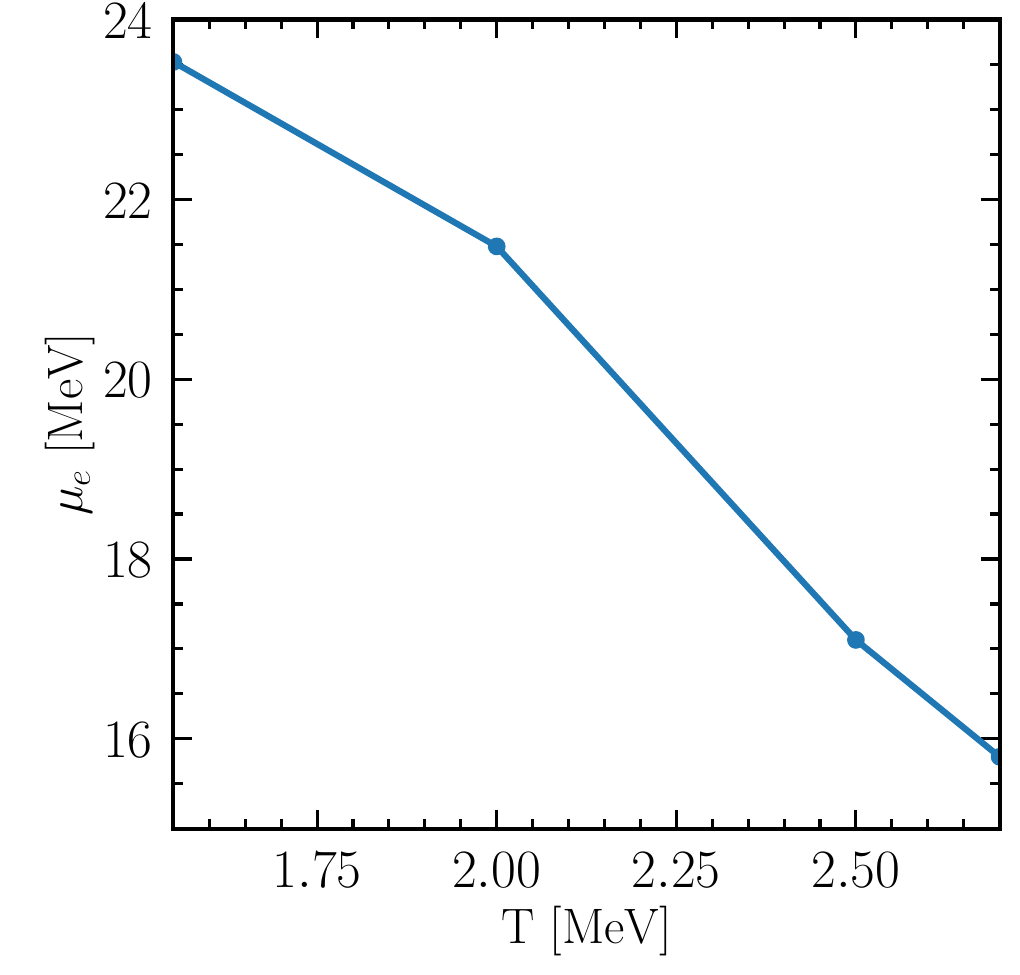} 
\includegraphics[width=0.45\columnwidth]{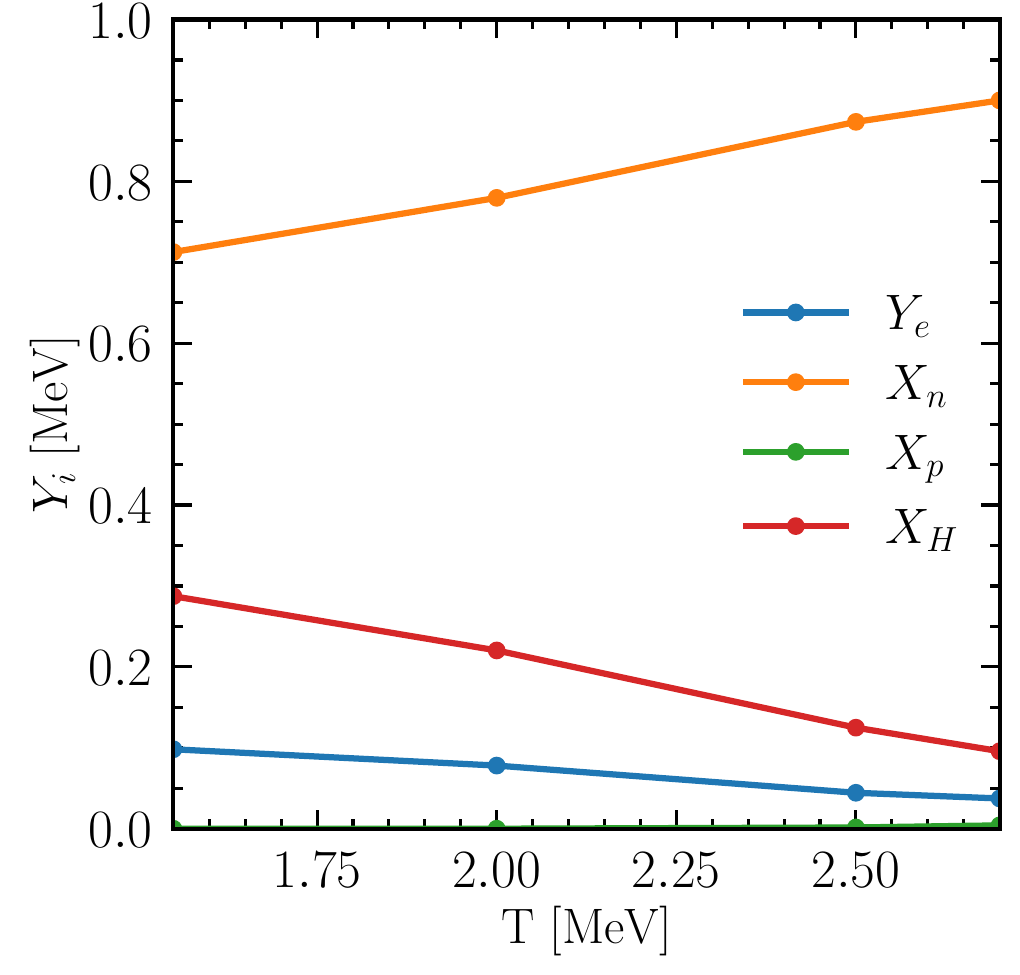} 
\caption{Solutions for the thermal and chemical equilibrium values of $\mu_e$, $Y_e$, $Y_p$, $Y_n$, $X_H$ for a given temperature of matter $T$ and $\mu_{\nu_e} = \mu_{\bar\nu_e} = 0$ in the presence of nuclei for considered in our work approximated EOS prescription.}
\label{Fig8}
\end{figure}

We have also solved for the new beta equilibrium $\mu_e = \delta m_{np} + \hat\mu$ for our approximate EOS, when the temperature of matter does not heat up too much, and the presence of nuclei is still allowed.  
Figure~\ref{Fig8} shows the solution to the new beta equilibrium with nuclei. The found solution indicates that regardless of the final temperature of matter, the new equilibrium should always have significantly lower $Y_e$ and $X_H$ than in the standard case. 

\section{Impact of the $\alpha$ particles on the New Weak Equilibrium with LNV $\nu$SI}

We have tested whether the results of our simulations change when the EOS used incorporates $\alpha$ particles. To do so, we repeated the simulation presented in Fig.~2 in the main text using the Lattimer and Swesty EOS~\cite{Lattimer:1991nc}.

\begin{figure}[t]
\includegraphics[width=0.32\columnwidth]{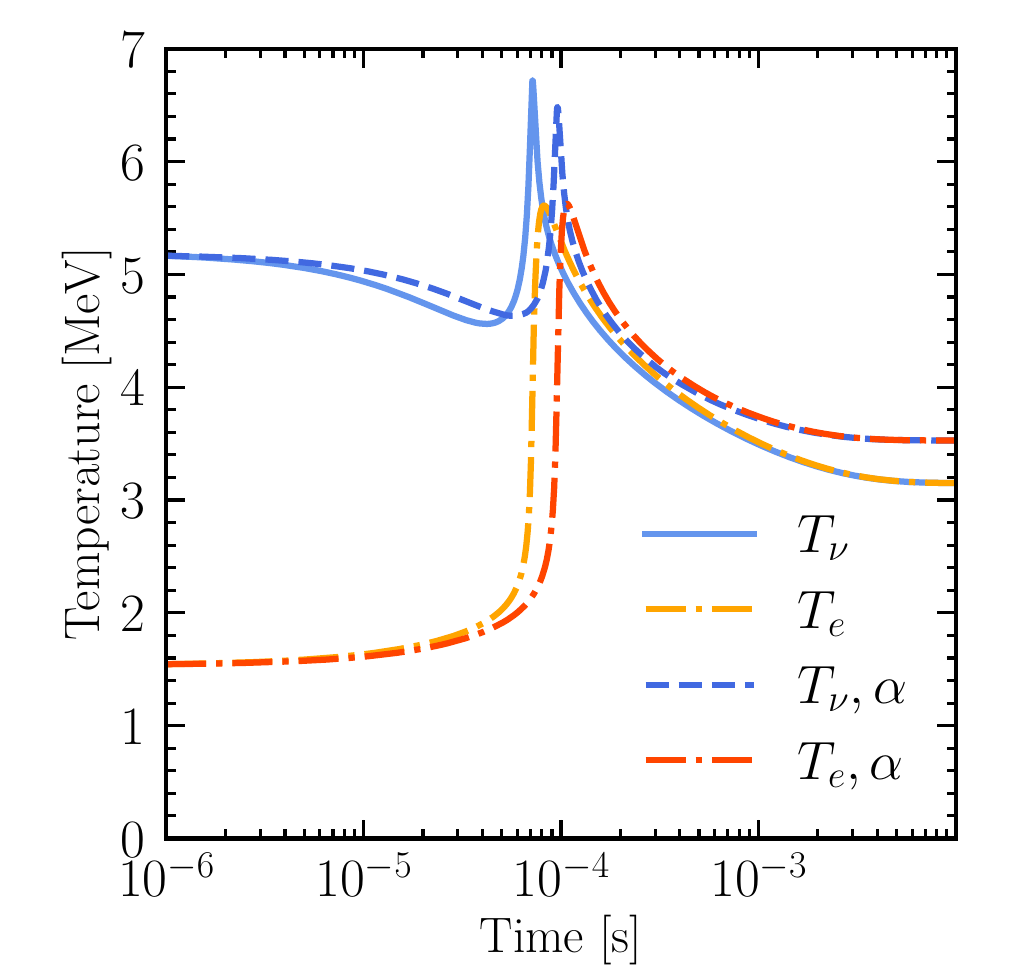} 
\includegraphics[width=0.32\columnwidth]{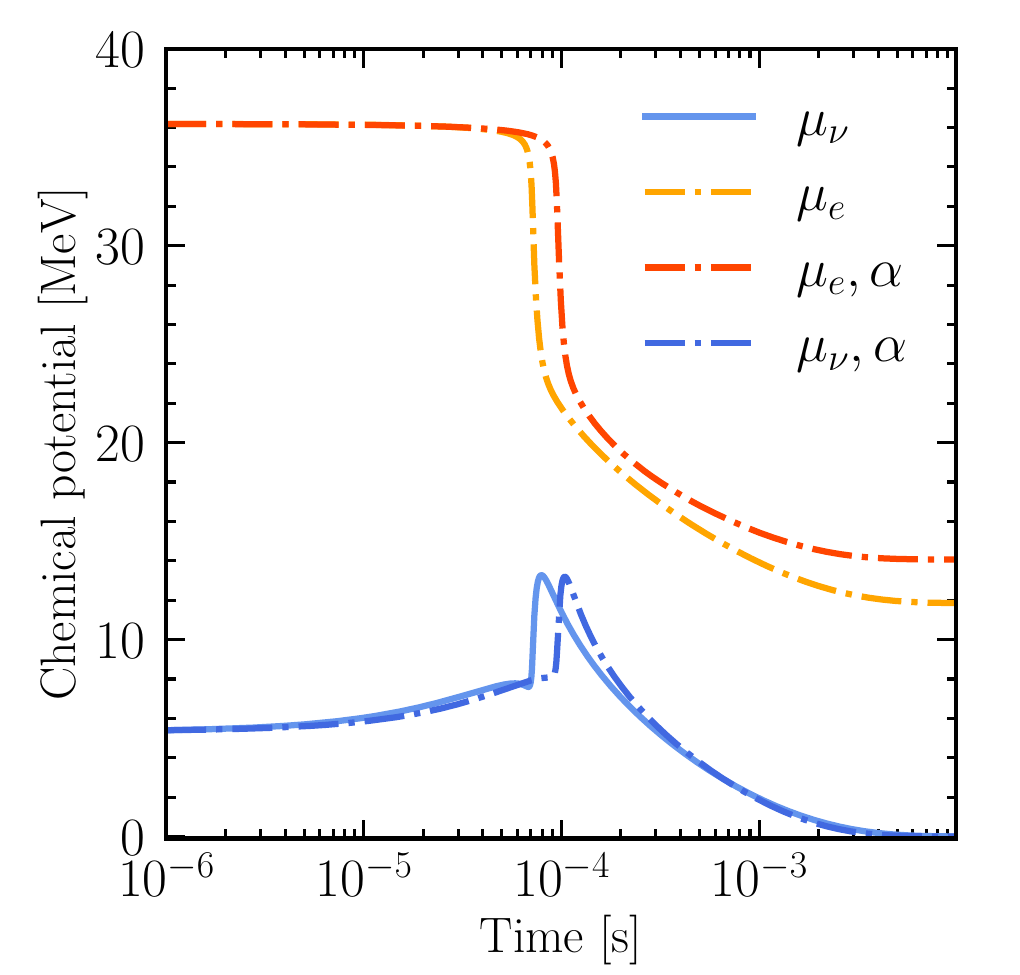} 
\includegraphics[width=0.32\columnwidth]{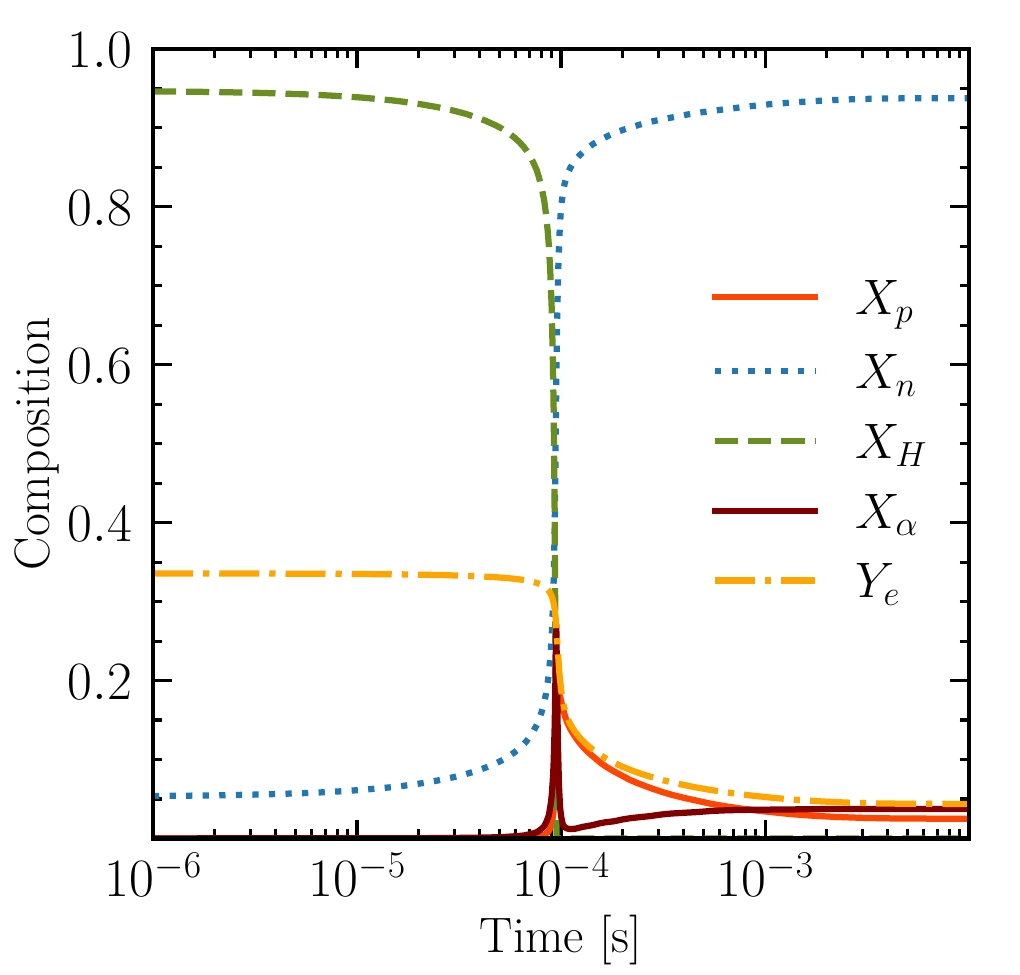} 
\caption{Temporal evolution of the effective temperatures (in MeV) for neutrinos ($T_{\nu}$), and the matter component ($T_e$) without $\alpha$ particles and with ($T_{\nu_i}, \alpha$ and $T_e,\alpha$) [\emph{left panel}]; electron and neutrino chemical potentials without $\alpha$ particles ($\mu_{\nu}$ and $\mu_{e}$) and with ($\mu_{\nu},\alpha$ and $\mu_{e},\alpha$) [\emph{middle panel}]; the composition of matter: mass fractions for free protons, $X_p$, free neutrons, $X_n$, heavy nuclei, $X_H$, alpha particles $\alpha$, and electron fraction $Y_e$ [\emph{right panel}].}
\label{Fig:9}
\end{figure}

Figure~\ref{Fig:9} compares the temporal evolution of neutrino and electron temperatures (\emph{left panel}) and chemical potentials (\emph{middle panel}) for simulations using an EOS that does not include $\alpha$ particles~\cite{Bethe:1979zd, Fuller:1981mu} and one that does~\cite{Lattimer:1991nc}. The results of both simulations are in qualitative agreement.
The right panel of Fig.~\ref{Fig:9} illustrates the temporal changes in the composition of matter for the simulation with $\alpha$ particles. The abundance of $\alpha$ particles is significant when the entropy/matter temperature rises, but as $Y_e$ decreases, their abundance decreases as well.
In general terms, as entropy increases as a result of the LNV $\nu$SI, the $\alpha$ mass fraction also rises. This occurs because, in nuclear statistical equilibrium (NSE), $\alpha$ particles become favored by the balance between disorder (higher entropy) and their nuclear binding energy. However, as weak equilibrium shifts and the matter becomes more neutron-rich (lower $Y_e$), $\alpha$ particles become less favored, as they contain equal numbers of neutrons and protons.

\end{document}